\documentclass[letterpaper,journal,twocolumn]{IEEEtran}

\ifCLASSINFOpdf
\else
\fi

\usepackage[cmex10]{amsmath}
\usepackage{amssymb}
\usepackage{mathtools}
\usepackage{booktabs}
\usepackage{multirow}
\usepackage{blindtext}
\usepackage{algorithm}
\usepackage{algorithmic}
\usepackage{color}
\usepackage{amsmath}
\usepackage{graphicx}
\usepackage{epstopdf}
\usepackage{pgfplots}
\pgfmathdeclarefunction{Krone}{1}{\pgfmathparse{or(abs(#1) > 0,0)}}
\pgfplotsset{major grid style={dotted}}
\usepackage{bbm}
\usepackage{relsize}

\usepackage[noadjust]{cite}
\newsavebox{\ieeealgbox}

\usepackage{array}

\usepackage{mdwmath}
\usepackage{mdwtab}

\usepackage{stfloats}
\fnbelowfloat

\newcommand{\Ab}{\mathbf{A}}
\newcommand{\Abta}{\Ab_{\tau}}
\newcommand{\Abtac}{\Ab_{\tau^{c}}}
\newcommand{\ab}{\mathbf{a}}
\newcommand{\Bb}{\mathbf{B}}

\newcommand{\Ib}{\mathbf{I}}
\newcommand{\Pb}{\mathbf{P}}

\newcommand{\Mb}{\mathbf{M}}

\newcommand{\bb}{\mathbf{b}}

\newcommand{\xb}{\mathbf{x}}

\newcommand{\rb}{\mathbf{r}}
\newcommand{\sbb}{\mathbf{s}}
\newcommand{\oneb}{\mathbf{1}}
\newcommand{\zerb}{\mathbf{0}}
\newcommand{\xbs}{\xb^*}
\newcommand{\xbh}{\widehat{\xb}}

\newcommand{\ub}{\mathbf{u}}

\newcommand{\db}{\mathbf{d}}
\newcommand{\eb}{\mathbf{e}}
\newcommand{\yb}{\mathbf{y}}
\newcommand{\ybt}{\widetilde{\yb}}

\newcommand{\zb}{\mathbf{z}}
\newcommand{\zbs}{\zb^*}
\newcommand{\zbt}{\widetilde{\zb}}
\newcommand{\zt}{\widetilde{z}}
\newcommand{\zbtta}{\zbt_{\tau}}
\newcommand{\zbttac}{\zbt_{\tau^c}}

\newcommand{\zbh}{\widehat{\zb}}
\newcommand{\zbhs}{\zbh_{\sigma}}

\newcommand{\zbhta}{\zbh_{\tau}}

\newcommand{\zbsta}{\zbs_{\tau}}

\newcommand{\zbta}{\zb_{\tau}}

\newcommand{\ubt}{\widetilde{\ub}}
\newcommand{\ubtta}{\ubt_{\tau}}
\newcommand{\ubttac}{\ubt_{\tau^{c}}}
\newcommand{\ubta}{\ub_{\tau}}

\newcommand{\wb}{\mathbf{w}}
\newcommand{\wbt}{\widetilde{\wb}}

\newcommand{\tb}{\mathbf{t}}
\newcommand{\tbh}{\widehat{\tb}}
\newcommand{\wt}{\widetilde{w}}
\newcommand{\Yb}{\mathbf{Y}}
\newcommand{\Zb}{\mathbf{Z}}

\newcommand{\Rbb}{\mathbb{R}}

\newcommand{\fsig}{f_{\sigma}}

\newcommand{\lamsig}{\lambda_{\sigma}}

\DeclareMathOperator{\diag}{diag}

\DeclareMathOperator*{\argmin}{argmin}

\DeclareMathOperator{\sign}{sgn}

\DeclareMathOperator{\supp}{supp}

\begin{document}


\title{A Class of Nonconvex Penalties Preserving Overall Convexity in Optimization-Based Mean Filtering}


%
%
%

\author{Mohammadreza~Malek-Mohammadi\textsuperscript{*}, Cristian R.~Rojas, \IEEEmembership{Member, IEEE}, Bo Wahlberg, \IEEEmembership{Fellow, IEEE}
\thanks{This work was supported by the Swedish Strategic Research Area ICT-TNG program and the Swedish Research Council (VR).}
\thanks{M.~Malek-Mohammadi, C.~R.~Rojas, and B.~Wahlberg are with the Department of Automatic Control and ACCESS Linnaeus Centre, KTH- Royal Institute of Technology, Stockholm, 10044, Sweden (e-mail: \{mohamma,crro,bo\}@kth.se).}}

\maketitle

\begin{abstract}
$\ell_1$ mean filtering is a conventional, optimization-based method to estimate the positions of jumps in a piecewise constant signal perturbed by additive noise. In this method, the $\ell_1$ norm penalizes sparsity of the first-order derivative of the signal. Theoretical results, however, show that in some situations, which can occur frequently in practice, even when the jump amplitudes tend to $\infty$, the conventional method identifies false change points. This issue is referred to as stair-casing problem and restricts practical importance of $\ell_1$ mean filtering. In this paper, sparsity is penalized more tightly than the $\ell_1$ norm by exploiting a certain class of nonconvex functions, while the strict convexity of the consequent optimization problem is preserved. This results in a higher performance in detecting change points. To theoretically justify the performance improvements over $\ell_1$ mean filtering, deterministic and stochastic sufficient conditions for exact change point recovery are derived. In particular, theoretical results show that in the stair-casing problem, our approach might be able to exclude the false change points, while $\ell_1$ mean filtering may fail. A number of numerical simulations assist to show superiority of our method over $\ell_1$ mean filtering and another state-of-the-art algorithm that promotes sparsity tighter than the $\ell_1$ norm. Specifically, it is shown that our approach can consistently detect change points when the jump amplitudes become sufficiently large, while the two other competitors cannot.
\end{abstract}



\begin{center} \bfseries
EDICS Category: DSP-RECO, SSP-SPRS, SSP-REST, SSP-FILT
\end{center}

\IEEEpeerreviewmaketitle

\section{Introduction}
\IEEEPARstart{E}{stimating} a piecewise constant (PWC) signal from noisy observations, usually referred to as mean filtering problem, has numerous applications in different areas of science and engineering. Applications of mean filtering include analysis of financial time series \cite{Tsay05} where one aims to recognize the time instants of changes in the trend of financial indicators (data), DNA segmentation \cite{ChenW09,KillFE12}, change point detection in biomedical engineering \cite{LaroNELP08}, health monitoring \cite{YangDA06}, network intrusion detection \cite{TartRBK06}, and total variation (TV) denoising in image processing \cite{RudiOF92,BeckT09} (see \cite{LittJ11,BassN93} for other lists of applications).

TV denoising, independently, is of central interest and can be utilized in many image processing tasks like computed tomography image reconstruction \cite{LiuLMLWZM14}, magnetic resonance image enhancement \cite{Keel03}, image segmentation \cite{KwonLW13}, and image and video denoising \cite{BeckT09}, among others. In a typical image, `edges' generally correspond to abrupt changes in the intensity level. These edges separate distinct regions from each other and occupy a small portion of the whole image area. Images can be, therefore, considered as two-dimensional PWC signals. TV denoising, to put it briefly, tries to reduce the noise from the flat regions while preserving the edges of the image.

Being piecewise constant implies that the number of changes occurring in the signal level is small when compared to the total number of samples. In other words, a change in the signal amplitude is a \emph{sparse} event in the history of the observations made from the signal. This sparsity indeed appears in the first-order derivative of the signal, and the tools available in the rich field of sparse signal processing can be employed to propose efficient algorithms. A well-known approach in this regard is $\ell_1$ mean filtering algorithm where the sparsity of the first-order derivative is penalized, in an optimization problem, by the $\ell_1$ norm. This algorithm leads to solving an unconstrained, strictly convex optimization problem, where the objective function is composed of the $\ell_1$ penalty and a data fidelity term. 
However, as shown in \cite{RojaW14}, in many cases, this approach is unable to precisely detect the change points (CP), the indexes in which there is a abrupt change in the amplitude of the PWC signal. Particularly, if two succeeding changes (jumps) are in the same direction---either increasing or decreasing in amplitude---$\ell_1$ mean filtering method usually finds false change points between the actual ones. This unfavorable effect is known as `stair-casing' problem \cite{GrasL10,RojaW15} and has been observed in TV-based denoising algorithms as gradual changes in flat regions of the recovered image \cite{CaseCN07,GrasL10}.


\subsection{Contribution}
The main purpose of the current paper is to propose an algorithm to enhance the possibility of CP recovery using a convex optimization program. Particularly, we would like to decrease the rate of identifying false changes, while the tractability of the resulting optimization problem as well as the possibility of theoretically supporting the algorithm is maintained. To be specific, we use a class of nonconvex penalties that approximates the $\ell_0$ norm more accurately than the $\ell_1$ norm yet preserves the overall convexity of the resulting optimization problem. We provide a sufficient condition for strict convexity of the proposed optimization problem that comes as no surprise to restrict the achievable accuracy of approximating the $\ell_0$ norm. This accuracy, however, is still better than that of the $\ell_1$ norm. Having this convexity, a computationally efficient algorithm for solving the optimization problem as well as a guarantee for convergence to the global minimizer is introduced. We also state a deterministic and a stochastic sufficient condition for exact change point recovery. These conditions have similarities to the well-known irrepresentable condition \cite{ZhaoY06,CandP09,Wain09} for the lasso estimator \cite{Tibs96}.

Our approach is inspired by the penalty functions used in \cite{MaleBS14,MaleKBJR15} to approximate sparsity in the context of compressed sensing (CS). However, in the underdetermined setting of CS, it is not possible to have a convex program if sparsity is promoted by a nonconvex penalty. As a result, in \cite{MaleBS14,MaleKBJR15}, the authors propose to use a continuation approach to decline the risk of getting trapped in a local solution without providing any theoretical guarantee. Here, we derive a condition for strict convexity, and this strict convexity allows us to guarantee convergence to the unique optimal solution. More remarkably, we are able to establish theoretical results for the performance of our algorithm.

In line with our idea, \cite{SelePB15} has also proposed some penalties that induce sparsity tighter than the $\ell_1$ norm; however, our algorithm has the following distinctions and contributions in comparison to the work of \cite{SelePB15}.

\begin{enumerate}
  \item We use a different class of penalties leading to a better performance in terms of CP recovery.
  \item We prove a sharper sufficient condition than the one stated in \cite{SelePB15} for convexity of the consequent optimization problem.
  \item Deterministic and stochastic conditions for CP recovery are established which are weaker than those of $\ell_1$ mean filtering.
  \item An efficient optimization method with a guaranteed convergence is suggested which is based on the weighted taut-string method of \cite{BarbS14}. 
  \item Numerical experiments demonstrate the superiority of our method over the algorithm of \cite{SelePB15} and $\ell_1$ mean filtering.
\end{enumerate}

\subsection{Notations and Outline}

\emph{Notations}: 
$\sign(x) = |x| / x$ for $x \neq 0$, and $\sign(0) = 0$. All functions and inequalities act component-wise for vector variables. $\zerb$ and $\oneb$ denote column vectors of zeros and ones of appropriate length, respectively. For a vector $\xb$, $\| \xb \|_1$, $\| \xb \|$, and $\| \xb \|_{\infty}$ denote the $\ell_1$, $\ell_2$, and $\ell_{\infty}$ norms, respectively. Further, $x_i$ and $[\xb]_i$ represent the $i$th element of $\xb$, $\xb_{I}$ denotes the subvector obtained from $\xb$ by keeping components indexed by the set $I$, and $\supp(\xb) = \{i|x_i \neq 0\}$ designates the support set of $\xb$. 
Also, $\xb \odot \yb$ indicates component-wise multiplication. $\Ib$ designates the identity matrix, and $\Ab_{S}$ represents the submatrix of $\Ab$ obtained by keeping those columns indexed by the set $S$. For matrices, $\| \Ab \|_{\infty}$ represents the matrix norm induced by the vector $\ell_{\infty}$ norm. $|S|$ designates the cardinality of the set $S$.

The rest of this paper is structured as follows. Section \ref{sec:AlgoDesc} describes the main idea of our algorithm and introduces a convexity condition for the optimization problem used in our algorithm as well as an efficient method for solving it. Section \ref{sec:TA} provides theoretical guarantees, while Section \ref{sec:NS} numerically shows the effectiveness of our algorithm. Although Section \ref{sec:C} concludes the paper, the proofs of all theoretical results are collected in the appendix.

\section{Description of the Algorithm} \label{sec:AlgoDesc}

Suppose that a set of samples, $y_1,\cdots,y_n$, has been observed, and they are collected in a vector $\yb = (y_1, \cdots,y_n)^T \in \Rbb^{n}$. These measurements are assumed to be generated according to the model
\begin{equation*}
\yb = \xbs + \wb,
\end{equation*}
where $\xbs$ is the unknown PWC signal, $\wb = (w_1, \cdots,w_n)^T$, and the $w_i$'s are independent and identically distributed Gaussian noise with zero mean and variance $\sigma_w^2$. In the mean filtering problem, the goal is to estimate $\xbs$ from the measurements vector $\yb$. Since 
$\xbs$ is piecewise constant, this a priori known structure should be employed to increase the accuracy of the estimation. However, this structure can be interpreted in different ways which leads to various algorithms; see \cite{BassN93,LittJ11} for a comprehensive review. 
To have a concise presentation, we restrict ourselves herein to optimization-based algorithms for recovering $\xbs$ from $\yb$.

\subsection{Motivation}
One way to exploit the structure of $\xbs$ is to penalize the number of changes occurring in its elements. More precisely, since $\xbs$ is PWC, its first-order (discrete) derivative is a sparse vector, and one can penalize this sparsity to obtain good estimations. An innate approach to induce sparsity is to use the $\ell_0$ norm, defined as $\| \xb \|_0 = \sum_i (1 - \delta(x_i))$, where $\delta(\cdot)$ designates the Kronecker delta function. Accordingly, using the $\ell_0$ and $\ell_2$ norms as sparsity and goodness-of-fit measures, it is possible to arrive at the optimization problem
\begin{equation} \label{l0min}
\min_{\xb \in \Rbb^{n}} \frac{1}{2n} \| \yb - \xb \|^2 + \lambda' \sum_{i = 1}^{n-1} 1 - \delta( x_{i+1} - x_i ),
\end{equation}
where $\lambda' > 0$ is a regularization parameter that balances between consistency to the measurements and sparsity of the first-order derivative of $\xb$. Any solution to \eqref{l0min} provides an estimation of $\xbs$ as a function of $\lambda'$. Nevertheless, \eqref{l0min} is, in essence, combinatorial and intractable for large values of $n$.

To have a computationally tractable optimization problem, the $\ell_0$ norm is replaced by its tightest convex relaxation leading to the well-known $\ell_1$ mean filtering program \cite{KimKBG09}
\begin{equation} \label{l1min}
\min_{\xb \in \Rbb^{n}} \frac{1}{2n} \| \yb - \xb \|^2 + \lambda \sum_{i = 1}^{n-1} | x_{i+1} - x_i|,
\end{equation}
where $\lambda > 0$, as $\lambda'$ in \eqref{l0min}, is a parameter that trades consistency to the measurements and sparsity of the first-order derivative of $\xb$. Although the $\ell_1$ norm is convex, it is quite well-known that this norm, when promotes sparsity, does not provide a performance close to that of the $\ell_0$ norm. The performance gap can be seen, for example, in the bias, support recovery, and estimation error \cite{Zou06,Zhan10,Zhan13,FanL01}. On the other hand, a number of theoretical and experimental results in compressive sensing and low-rank matrix recovery (LMR) frameworks suggests that better approximations of the $\ell_0$ norm and the matrix rank result in better performances \cite{Zhan10,CandWB08,Char07,WuC13,MaleBS14,MaleKBJR15,MaleBS15,MaleBAJ14}. These studies inspire the same result in the mean filtering problem. In other words, we expect that a more accurate approximation of the $\ell_0$ norm as a sparsity measure will give rise to a better performance in recovering $\xbs$. Below, we pursue this idea to propose a new algorithm for the mean filtering problem, while theoretical justification of a higher performance is deferred to Section \ref{sec:TA}.

\subsection{The Core Idea} \label{sec:Core}
The main idea of our algorithm is to exploit a class of nonconvex functions to approximate the $\ell_0$ norm more accurately than that of the $\ell_1$ norm. We use the class of nonconvex penalties introduced in \cite{MaleBS14,MaleKBJR15} for CS and LMR problems. Nevertheless, contrary to \cite{MaleBS14,MaleKBJR15}, the resulting optimization problem is convex. In fact, although the penalty function is nonconvex, since it has a scaling parameter that controls the degree of nonconvexity and is put beside the strictly convex term, $\| \yb - \xb \|^2$, 
it is possible to keep the optimization problem convex. This, as will be shown later, is realized by choosing the scaling parameter in an appropriate way.

Recalling that $\| \xb \|_0 = \sum_i (1 - \delta(x_i))$, approximation of the $\ell_0$ norm can be simplified to the task of approximating $1 - \delta(x)$. This can be realized with the following class of one-variable, nonconvex functions.

\smallskip

\newtheorem{Prop1}{Property}
\begin{Prop1} \label{Prop1}
Let  $f: \Rbb \rightarrow [-\infty,\infty)$ and define $\fsig(x) \triangleq f(\frac{x}{\sigma})$ for any $\sigma > 0$. The function $f$ possesses Property \ref{Prop1}, if
\begin{enumerate}
\item[(a)] $f$ is real analytic on $(x_0,\infty)$ for some $x_0 < 0$,

\item[(b)] $\forall x \geq 0$, $f''(x) \geq -\mu$, where $\mu  > 0$ is some constant,

\item[(c)] $f$ is concave on $\Rbb$,

\item[(d)] $f(x) = 0 \Leftrightarrow x = 0$ and $f'(0) = 1$,

\item[(e)] $\lim_{x\rightarrow +\infty}f(x) = 1$.


\end{enumerate}
\end{Prop1}

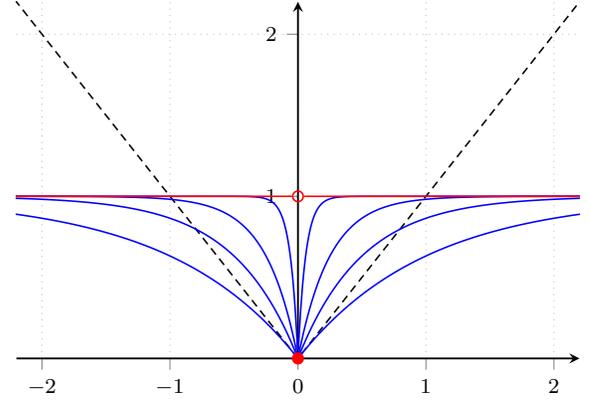
\begin{figure}[tb]
\centering
\pgfplotsset{every axis/.append style={font = \footnotesize, line width=0.7pt}}
\begin{tikzpicture}
	\begin{axis}[
	width = 0.50\textwidth,
	height = 180pt,
	ymin = 0, ymax = 2.2,
	xmin = -2.2, xmax = 2.2,
	ytick={1,2},
	xtick={-2,-1,0,1,2},
	axis x line=bottom,
	axis y line=center,
	scaled ticks = false,
	tick align=outside,
	xmajorgrids = true,
	ymajorgrids = true,
	]
	\addplot[black,densely dashed,domain=-3.75:3.75,samples = 1000,line width = 0.6] (\x,{abs(\x)});
    \addplot[blue,domain=-3.75:3.75,samples = 1000,line width = 0.6] (\x,{1 - exp(-abs(\x)/1)});
    \addplot[blue,domain=-3.75:3.75,samples = 1000,line width = 0.6] (\x,{1 - exp(-abs(\x)/0.5)});
    \addplot[blue,domain=-3.75:3.75,samples = 1000,line width = 0.6] (\x,{1 - exp(-abs(\x)/0.25)});
    \addplot[blue,domain=-3.75:3.75,samples = 1000,line width = 0.6] (\x,{1 - exp(-abs(\x)/0.06)});
	\addplot[red,domain=-3.75:-0.05,samples = 500,line width = 0.6] (\x,{Krone(\x)});
	\addplot[red,domain=0.05:3.75,samples = 500,line width = 0.6] (\x,{Krone(\x)});
	\addplot+[red,mark=o,line width = 0.6] coordinates {(0,1)};
	\addplot[red,mark color=red,mark=*,line width = 0.6] coordinates {(0,0)};
	\end{axis}
\end{tikzpicture}
\caption{$\fsig(|x|) = 1 - e^{-|x|/\sigma}$ provides better approximations of $1 - \delta(x)$ when $\sigma$ decreases. In this plot, the blue curves show $\fsig(|x|)$ for a decreasing sequence of $\sigma$'s, where for the one closely matches $1 - \delta(x)$, $\sigma$ equals 0.06. The black and red curves respectively show $|x|$ and $1 - \delta(x)$.} \label{fig:FuncPlot}
\end{figure}

Among the functions fulfilling Property \ref{Prop1} (see \cite{MaleBS14} for other examples), $f(x) = 1-e^{-x}$ is of central interest in the rest of this paper. Moreover, notice that the scaling parameter $\sigma$ reflects accuracy; the smaller $\sigma$, the better accuracy in approximating the $\ell_0$ norm. This can be seen in Fig.~\ref{fig:FuncPlot}, where $\fsig(|x|) = 1 - e^{-|x|/\sigma}$ for sufficiently small values of $\sigma$ provides a close fitting to $1 - \delta(x)$. 

Exploiting the class of nonconvex functions having Property \ref{Prop1}, the proposed optimization problem for estimating $\xbs$ is formulated as
\begin{equation} \label{fmin}
\min_{\xb \in \Rbb^{n}} \frac{1}{2n} \| \yb - \xb \|^2 + \lamsig \sum_{i = 1}^{n-1} \fsig( | x_{i+1} - x_i |),
\end{equation}
where $\lamsig > 0$ is a regularization parameter depending on $\sigma$. To get close to \eqref{l0min}, one needs to choose $\sigma$ as small as possible. However, it is not possible to have $\sigma$ arbitrarily small as for $\sigma = 0$, program \eqref{fmin} is not tractable.
It is also insightful to look at the asymptotic behaviours of $\fsig(|x|)$. It can be seen that $\fsig(|x|)$ converges pointwise to $1 - \delta(x)$ when $\sigma$ tends to 0; that is,
\begin{equation*}
\lim_{\sigma \to 0^{+}} \fsig(|x|)= \left\{
	\begin{IEEEeqnarraybox}[][c]{l?s}
    \IEEEstrut
	0 & if  $x = 0$\\
	1 & otherwise.
	\IEEEstrut
    \end{IEEEeqnarraybox}
    \right.
\end{equation*}
This means that to have a good approximation, we should choose $\sigma$ as small as possible. In addition, we will prove later that $\sigma \fsig(|x|)$ converges to $|x|$ when $\sigma \to \infty$. In the light of these facts, it is possible to conclude that the class of functions possessing Property \ref{Prop1} is interpolating between the $\ell_0$ and $\ell_1$ norms. It is not surprising, hence, to expect a performance better than that of the $\ell_1$ norm.

The obstacle that prevents $\sigma$ from being arbitrarily small is strict convexity of \eqref{fmin}. This is mathematically characterized in terms of the scaling parameter $\sigma$ in Theorem \ref{ConvexityThm}, Section \ref{sec:CC}, yet a rationale is as follows. As the first term in the objective function of \eqref{fmin} is strictly convex, there is some room for the second term to be nonconvex to preserve strict convexity of the whole objective function. Obviously, when $\sigma$ increases $\sum_{i = 1}^{n-1} \fsig( | x_{i+1} - x_i |)$ tends to become a convex function, while for small $\sigma$ the degree of nonconvexity increases. As a result, we need to restrict the value of $\sigma$ from below in order to have a convex program.

\subsection{Relation to $\ell_0$ Minimization}
Before stating the convexity condition and the derivation of the optimization method for solving \eqref{fmin} in Sections \ref{sec:CC} and \ref{sec:OM}, it is quite useful to look at the intuition behind the final optimization program which should be solved iteratively. In fact, the following explanation completes the motivation presented in Section \ref{sec:Core} about approximating the Kronecker delta function. Let $\xbh^{(k)}$ denote the solution at the $k$th iteration,  
the next solution in the sequence of the minimizers converging to the optimal solution of \eqref{fmin} is obtained as
\ifCLASSOPTIONjournal
\begin{multline} \label{FinProgram}
\xbh^{(k+1)} = \argmin_{\xb} \Big \{ \frac{1}{2n} \| \yb - \xb \|^2 + \\
\frac{\lamsig}{\sigma} \sum_{i = 1}^{n - 1} f'(\frac{|\widehat{x}_{i+1}^{(k)} - \widehat{x}_{i}^{(k)}|}{\sigma}) |x_{i+1} - x_{i}| \Big \},
\end{multline}
\else
\begin{equation} \label{FinProgram}
\xbh^{(k+1)} = \argmin_{\xb} \Big \{ \frac{1}{2n} \| \yb - \xb \|^2 + \frac{\lamsig}{\sigma} \sum_{i = 1}^{n - 1} f'(\frac{|\widehat{x}_{i+1}^{(k)} - \widehat{x}_{i}^{(k)}|}{\sigma}) \big |x_{i+1} - x_{i} \big| \Big \},
\end{equation}
\fi
where $f'(|\widehat{x}_{i+1}^{(k)} - \widehat{x}_{i}^{(k)}|/\sigma)$ represents the derivative of $f(\cdot)$ calculated\footnote{It is worth emphasizing that $f(|\cdot|)$ is not differentiable, while $f(\cdot)$ is.} at point $|\widehat{x}_{i+1}^{(k)} - \widehat{x}_{i}^{(k)}| / \sigma$. The above program is a (re)weighted version of $\ell_1$ mean filtering program \eqref{l1min}, where the weights depend on the previous solution as well as the selected approximating function.

For the sake of simplicity of explanation, let us focus on $f(x) = 1 - e^{-x}$. With this choice, the $i$th weight is $\exp\big(- |\widehat{x}_{i+1}^{(k)} - \widehat{x}_{i}^{(k)}| / \sigma\big)$. Moreover, as will be explained, $\lamsig$ should be equal to $\lambda \sigma$, where $\lambda$ is the regularization parameter used in \eqref{l1min}. Altogether, \eqref{FinProgram} converts to
\ifCLASSOPTIONjournal
\begin{multline} \label{FinProgramExp}
\xbh^{(k+1)} = \argmin_{\xb} \Big \{ \frac{1}{2n} \| \yb - \xb \|^2 + \\
\lambda \sum_{i = 1}^{n - 1} e^{-|\widehat{x}_{i+1}^{(k)} - \widehat{x}_{i}^{(k)}| / \sigma} |x_{i+1} - x_{i}| \Big \}.
\end{multline}
\else
\begin{equation} \label{FinProgramExp}
\xbh^{(k+1)} = \argmin_{\xb} \Big \{ \frac{1}{2n} \| \yb - \xb \|^2 + \lambda \sum_{i = 1}^{n - 1} e^{-|\widehat{x}_{i+1}^{(k)} - \widehat{x}_{i}^{(k)}| / \sigma} |x_{i+1} - x_{i}| \Big \}.
\end{equation}
\fi
Program \eqref{l1min} penalizes a nonzero $(x_{i+1} - x_{i})$ with its absolute value which results in the shrinkage of the amplitudes of the estimated solution. In fact, due to a higher penalty for a larger amplitude of $(x_{i+1} - x_{i})$, \eqref{l1min} tends to underestimate $\xbs$. In contrast, \eqref{l0min} penalizes all nonzero components $(x_{i+1} - x_{i})$ equally, and \eqref{FinProgramExp} mimics this behaviour of \eqref{l0min}. More specifically, when $(\widehat{x}_{i+1}^{(k)} - \widehat{x}_{i}^{(k)})$ is nonzero in the previous iteration, $|x_{i+1} - x_{i}|$ at the $(k+1)$th iteration has a weight strictly smaller than 1; i.e., the $i$th component of the penalty is weighted as
\begin{equation*}
|x_{i+1} - x_{i}| / \exp\big( |\widehat{x}_{i+1}^{(k)} - \widehat{x}_{i}^{(k)}| / \sigma \big).
\end{equation*}
This shows that the penalty associated with $(x_{i+1} - x_{i})$ decreases at the $(k+1)$th iteration, if $(\widehat{x}_{i+1}^{(k)} - \widehat{x}_{i}^{(k)})$ increases. Consequently, \eqref{FinProgramExp} tends to have a behaviour closer to that of the original optimization problem \eqref{l0min}.

One might argue that a weight proportional to $1 / |\widehat{x}_{i+1}^{(k)} - \widehat{x}_{i}^{(k)}|$ in the above expression will lead to a performance similar to that of \eqref{l0min} because when the optimization algorithm converges $\xbh^{(k + 1)}$ will coincide to $\xbh^{(k)}$. Nevertheless, a reasoning to oppose this opinion is as follows. It is well known that this kind of weight arises from approximating $1 - \delta(x)$ with $\log(|x|)$ \cite{FazeHB03,SelePB15}. However, another important point is a constant that appears in the numerator of the weight. 
This constant is dictated by the convexity condition, and if it is not equal to 1, then the penalization will differ from that of \eqref{l0min}. This has been also observed in \cite{SelePB15}, where another approximating function outperforms the $\log$ function.

\subsection{The Convexity Condition} \label{sec:CC}
To provide a sufficient condition for strict convexity of \eqref{fmin}, we need to introduce an 
equivalent form of \eqref{fmin} in the following proposition. The proof of the proposition follows directly from \cite[Lem.~2.4]{RojaW14}.

\newtheorem{Pro1}{Proposition}
\begin{Pro1} \label{LassoEqv}
Program \eqref{fmin} is equivalent to
\begin{equation} \label{fminlasso}
\min_{\zb \in \Rbb^{n - 1}} \frac{1}{2n} \| \ybt - \Ab \zb \|^2 + \lamsig \sum_{i = 1}^{n-1} \fsig( | z_i |),
\end{equation}
where 
$\widetilde{y}_i = y_i - \frac{1}{n}\sum_{j = 1}^{n} y_j,~1 \leq i \leq n,$ and $\Ab \in \Rbb^{n \times n-1}$ is given by
\begin{equation} \label{ADef}
[\Ab]_{i,j} = \left\{
\begin{IEEEeqnarraybox}[][c]{l?s}
\IEEEstrut
\frac{j - n}{n} & $i \leq j$ \\
\frac{j}{n} & $i > j$
\IEEEstrut
\end{IEEEeqnarraybox}
\right..
\end{equation}
Moreover, if $\zbh$ denotes a solution to \eqref{fminlasso}, then the associated solution to \eqref{fmin} can be obtained by
\begin{equation*}
\widehat{x}_1 = \frac{1}{n} \sum_{i=1}^{n} y_i - \frac{1}{n} \sum_{i=1}^{n - 1} \sum_{j=1}^{i} \widehat{z}_j, ~ \widehat{x}_i = \widehat{x}_1 + \sum_{j=1}^{i-1} \widehat{z}_j, \, 2 \leq i \leq n.
\end{equation*}
\end{Pro1}

\smallskip

Proposition \ref{LassoEqv} suggests an equivalent form for program \eqref{fmin} in which the penalty term, $\sum_{i = 1}^{n-1} \fsig( | z_i |)$, is separable with respect to the components of $\zb$. \footnote{Recall that in \eqref{fmin}, the $i$th component of the penalty term, $\fsig(|x_{i + 1} - x_i |)$, depends on both $x_{i + 1}$ and $x_{i}$; hence, the penalty, $\sum_{i = 1}^{n-1} \fsig( | x_{i+1} - x_i |)$, is not separable.} This equivalent form considerably shortens the mathematical manipulation needed to prove the theoretical analyses presented in this paper. To put the observation model in accordance to the above equivalent form, 
it is necessary to update the model to
\begin{equation}\label{lassoSigMod}
\ybt = \Ab \zbs + \wbt,
\end{equation}
where
\begin{equation*}
\widetilde{w}_i = \frac{n-1}{n} w_i - \frac{1}{n} \sum_{j \neq i} w_j, \quad 1 \leq i \leq n,
\end{equation*}
and $z_i^{*} = x_{i+1}^* - x_{i}^*,~1 \leq i \leq n - 1$ \cite[Lem.~2.4]{RojaW14}.



Now, we are ready to state one of the key results of this paper in the following theorem. This theorem provides a condition for strict convexity of \eqref{fmin} and hence the uniqueness of the solution. In addition, all the theoretical guarantees established in Section \ref{sec:TA} rely on the strict convexity of \eqref{fmin}. As a side result, it will also ease showing the convergence of \eqref{FinProgram} to the global minimizer of \eqref{fmin}.

\newtheorem{Thm1}{Theorem}
\begin{Thm1} \label{ConvexityThm}
The cost function in \eqref{fmin} is strictly convex provided that
\begin{equation} \label{muCond}
\sigma^2 \geq \frac{\lamsig n}{s_{\min}} \mu,
\end{equation}
where $s_{\min}$ denotes the smallest eigenvalue of $\Ab^T \Ab$, $\Ab$ is defined in \eqref{ADef}, and $\mu$ is the constant defined in Property \ref{Prop1}-(b). Under the strict inequality in \eqref{muCond}, \eqref{fminlasso} is strictly convex too.
\end{Thm1}

\smallskip

The following remarks are in order.

\noindent \textbf{Remark 1.} The same kind of nonconvex penalties 
also appears in the framework of CS \cite{MaleKBJR15}. However, since the sensing matrix in that setting is not full column rank, it is not possible to have an overall convex program for any finite $\sigma$. 

\noindent \textbf{Remark 2.} Since $\Ab$ is fully determined for any $n$, $s_{\min}$ can be calculated numerically beforehand. Nevertheless, by running a simple simulation, it can be seen that $s_{\min}$ gradually decreases from $\frac{1}{2}$ to $\frac{1}{4}$, when $n$ goes from $2$ to very large values. This is in accordance with the result of \cite{SelePB15}, which when translated to our setting proves that $\sigma^2 > 4 \mu \lamsig n$ is a sufficient condition for strict convexity of \eqref{fmin}. In fact, if $s_{\min} \to \frac{1}{4}$ when $n \to \infty$, our condition in Theorem \ref{ConvexityThm} coincides with the convexity condition in \cite{SelePB15} showing that Theorem \ref{ConvexityThm} proves a strictly sharper condition.

\subsection{The Proposed Optimization Method for Solving \eqref{fmin}} \label{sec:OM}
To solve \eqref{fmin}, we use the majorization-minimization (MM) technique \cite{HuntL04}. To begin, \eqref{fminlasso} which is equivalent to \eqref{fmin} is converted to a program with a differentiable objective function in the following proposition whose proof easily follows from \cite[Thm.~1]{MaleKBJR15}. This is done by decoupling positive and negative entries of $\zb$.

\newtheorem{Pro2}[Pro1]{Proposition}
\begin{Pro1} \label{AlgoEqv}
Let $\tb = (\zb_p^T,\, \zb_n^T)^T$ denote a column vector of length $2n - 2$, where $\zb_p = \max(\zb, \zerb)$ and $\zb_n = -\min(\zb, \zerb)$. Let also $\Bb = [\Ab,-\Ab]$. \eqref{fminlasso} is equivalent to
\begin{equation} \label{fminsemilasso}
\min_{\tb \in \Rbb^{2n - 2}} \frac{1}{2n} \big \| \ybt - \Bb \tb \big \|^2 + \lamsig \sum_{i = 1}^{2n - 2} \fsig(t_i)~~~\text{s.t.}~~~
\tb \geq \zerb.
\end{equation}
\end{Pro1}

Since $|\cdot|$ is dropped from the argument of $\fsig$ in \eqref{fminlasso}, the objective function in \eqref{fminsemilasso} is now differentiable. Applying the first-order concavity condition for $\fsig(x)$ when $x \geq 0$ and neglecting the constant terms, the MM technique leads to iteratively solving
\begin{equation} \label{PreFinProgram1}
\tbh^{(k+1)} = \argmin_{\tb} \Big \{ \frac{1}{2n} \| \ybt - \Bb \tb \|^2 + \frac{\lamsig}{\sigma} \sum_{i = 1}^{2n - 2} f'(\frac{\widehat{t}_i^{(k)}}{\sigma}) t_i \Big | \tb \geq \zerb \Big \}
\end{equation}
until convergence. By applying Propositions \ref{AlgoEqv} and \ref{LassoEqv} to the above program, it can be converted back to a form similar to \eqref{fmin}. Namely, to solve \eqref{fmin}, we propose to solve
\ifCLASSOPTIONjournal \begin{multline*} 
\xbh^{(k+1)} = \argmin_{\xb} \Big \{ \frac{1}{2n} \| \yb - \xb \|^2 + \\
\frac{\lamsig}{\sigma} \sum_{i = 1}^{n - 1} f'(\frac{|\widehat{x}_{i+1}^{(k)} - \widehat{x}_{i}^{(k)}|}{\sigma}) |x_{i+1} - x_{i}| \Big \}
\end{multline*}
\else
\begin{equation*} 
\xbh^{(k+1)} = \argmin_{\xb} \Big \{ \frac{1}{2n} \| \yb - \xb \|^2 + \frac{\lamsig}{\sigma} \sum_{i = 1}^{n - 1} f'(\frac{|\widehat{x}_{i+1}^{(k)} - \widehat{x}_{i}^{(k)}|}{\sigma}) |x_{i+1} - x_{i}| \Big \}
\end{equation*}
\fi
iteratively until converging to a solution.

As discussed earlier, the above program is a weighted version of \eqref{l1min}. This program, thus, can be solved efficiently using the weighted taut-string algorithm of \cite{BarbS14}. This algorithm 
extends the taut-string algorithm of \cite{Cond13} which is originally designed for solving \eqref{l1min}. The worst-case complexity of the taut-string algorithm in \cite{Cond13} is of order $n^2$, while in practice, the complexity is close to order $n$. Consequently, the worst-case complexity of our approach might be of order $n^2 m$, where $m$ is the number of iterations needed for the convergence of \eqref{FinProgram}.

The following remarks describe other implementation details of the proposed optimization method.

\noindent \textbf{Remark 3.} Following the same line of argument as in \cite{MaleKBJR15}, it can be seen that a reasonable choice for $\lamsig$ as a function of $\sigma$ is $\lamsig = \lambda \sigma$, where $\lambda$ is the parameter used in \eqref{l1min}.

\noindent \textbf{Remark 4.} To initialize the sequence of optimization problems in \eqref{FinProgram}, one way is to start with $\xbh^{(0)} = \zerb$. The next point, $\xbh^{(1)}$, then will be equal to the solution of \eqref{l1min}. However, this choice can be motivated by the following proposition too.

\smallskip

\newtheorem{Pro3}[Pro1]{Proposition}
\begin{Pro3} \label{InftyPro}
Assume that $\lamsig = \lambda \sigma$, and let $\zbhs$ denote the unique solution to \eqref{fminlasso} for a given $\sigma > \lambda n \mu / s_{\min}$. Further, let
\begin{equation} \label{ConvEqv}
\zbt = \argmin_{\zb} \big \{ \frac{1}{2n} \| \ybt - \Ab \zb \|^2 + \lambda \| \zb \|_1 \big \}
\end{equation}
designate the solution corresponding to the $\ell_1$ mean filtering method. \footnote{It is shown in \cite{RojaW14} that \eqref{ConvEqv} is equivalent to \eqref{l1min}; see \cite[Lem.~2.4]{RojaW14} for further detail.} Then $\lim_{\sigma \to \infty} \zbhs = \zbt$.
\end{Pro3}

\smallskip

\noindent The above proposition shows that when $\sigma \to \infty$, which corresponds to the worst accuracy in approximating $1 - \delta(\cdot)$, solving \eqref{fmin} is equivalent to solving \eqref{l1min}. This is another indication that we should expect a better performance than that of $\ell_1$ mean filtering. 

\noindent \textbf{Remark 5.} If $\sigma$ is chosen large enough so that \eqref{fmin} is strictly convex, then \cite[Thm.~2.1]{SchiSW10} implies that
\begin{equation} \label{PreFinProg1}
\zbh^{(k+1)} = \argmin_{\zb} \Big \{ \frac{1}{2n} \| \ybt - \Ab \zb \|^2 + \frac{\lamsig}{\sigma} \sum_{i = 1}^{n - 1} f'(\frac{|\widehat{z}_{i}^{(k)}|}{\sigma}) |z_{i}| \Big \}
\end{equation}
converges to the unique minimizer of \eqref{fminlasso}. In fact, any function possessing Property \ref{Prop1} satisfies the regularity condition stated in \cite[Thm.~2.1]{SchiSW10} for the singular-at-the-origin penalties. Following the same line of arguments as in \cite[Lem.~2.4]{RojaW14} and Proposition \ref{LassoEqv}, it can be verified that \eqref{PreFinProg1} and \eqref{FinProgram} are equivalent. This shows the convergence of the sequence generated by \eqref{FinProgram} to the global minimizer of \eqref{fmin}. This result is summarized in the following proposition whose proof easily follows from \cite[Thm.~2.1]{SchiSW10}.

\newtheorem{Pro5}[Pro1]{Proposition}
\begin{Pro5} \label{Convergence}
Assume that $\sigma^2 > \frac{\lamsig n}{s_{\min}} \mu$. The sequence of minimizers generated by \eqref{FinProgram} is convergent to the global minimizer of \eqref{fmin}.
\end{Pro5}

\smallskip

Considering the above explanation, our proposed algorithm can be summarized in Algorithm \ref{alg:Algo}.

\begin{algorithm}[t] \small
\caption{The proposed algorithm}\label{alg:Algo}
\mbox{Input: $\yb, \lambda, \sigma$}\\
\mbox{Initialization:}
\begin{algorithmic}[1]
\STATE $\epsilon$: a stopping threshold.
\end{algorithmic}
\mbox{Body:}
\begin{algorithmic}[1]
\STATE $k = 0, \xbh^{(0)} = \zerb.$
\WHILE{$d > \epsilon$}
\STATE Find $\xbh^{(k+1)}$ in \eqref{FinProgram} using the weighted taut-string algorithm.
\STATE $d = \| \xbh^{(k + 1)}  - \xbh^{(k)} \| / \| \xbh^{(k)} \|.$
\STATE $k = k + 1.$
\ENDWHILE
\end{algorithmic}
\mbox{Output: $\xbh^{(k)}$}
\end{algorithm}

\section{Theoretical Analysis} \label{sec:TA}
The most important aspect of solving the mean filtering problem is to find the change points precisely. When they are recognized, it is possible to use an optimal estimator to improve the quality of the mean estimations. Having this in mind, we mainly focus on deriving performance guarantees for the change-point-recovery capability of our proposed algorithm in this section. In particular, a lemma is first stated that provides a sufficient condition for exact change point recovery. This lemma, however, guarantees the CP recovery given a realization of the noise vector. To extend this result to the case that the noise vector is drawn from a Gaussian distribution, an asymptotic setting is considered where $n \to \infty$. It will be shown that under a condition comparable to the irrepresentable condition \cite{ZhaoY06}, all CPs can be recovered by our algorithm with an overwhelming probability. Comparison to the associated conditions for $\ell_1$ mean filtering will follow afterwards.

It is always assumed in this section that $\lamsig = \lambda \sigma$ and 
$\sigma > \frac{\lambda n}{s_{\min}} \mu$ implying that optimization problems \eqref{fmin} and \eqref{fminlasso} are strictly convex. To derive the theoretical results, it is mainly focused on program \eqref{fminlasso}. This does not confine our analysis as \eqref{fminlasso} and \eqref{fmin} are equivalent, yet simplifies the derivations substantially. We start with the following basic lemma which characterizes optimality conditions for program \eqref{fminlasso}. The proof easily follows from the Karush-Kuhn-Tucker condition \cite{LuenY84}.

\newtheorem{Lem1}{Lemma}
\begin{Lem1} \label{OCLemma}
$\zbh$ is an optimal solution to \eqref{fminlasso} if and only if there exists a vector $\ub = (u_1,\cdots,u_{n-1})^T$ with elements $u_i \in \partial \fsig(|\widehat{z}_i|)$ such that
\begin{equation} \label{OC}
\frac{1}{n} \Ab^T ( \ybt -  \Ab \zbh ) = \lamsig \ub,
\end{equation}
where $\partial \fsig(|\widehat{z}_i|)$ denotes the Clarke subdifferential of $\fsig(|\cdot|)$ at $\widehat{z}_i$ \cite{Clar90} defined as
\begin{equation*}
\partial \fsig(|\widehat{z}_i|) = \left\{
\begin{IEEEeqnarraybox}[][c]{l?s}
\IEEEstrut
\Big\{ \frac{\sign(\widehat{z}_i)}{\sigma} f'(\frac{|\widehat{z}_i|}{\sigma}) \Big\} & if $\widehat{z}_i \neq 0$ \\
\big[-\frac{1}{\sigma},\frac{1}{\sigma}\big] & if $\widehat{z}_i = 0$.
\IEEEstrut
\end{IEEEeqnarraybox}
\right.
\end{equation*}
\end{Lem1}

\smallskip
To state the main lemma of this section, we need to introduce a restricted version of \eqref{fminlasso}. Let $\tau = \supp(\zbs)$ denote the support set of the true solution. 
We consider the following restricted program
\begin{equation} \label{fminlasso_res}
\min_{\zbta} \frac{1}{2n} \| \ybt - \Abta \zbta \|^2 + \lamsig \sum_{i} \fsig( | [\zbta]_i |)
\end{equation}
in our analysis. The above program is 
also strictly convex because \cite[Thm.~7.3.9]{HornJ90} implies that the smallest singular value of $\Ab_{\tau}$ is larger than that of $\Ab$. Thus, Theorem \ref{ConvexityThm} proves that \eqref{fminlasso_res} is strictly convex. The introduction of this restricted program, inspired by the work of \cite{CandP09,Trop06,Wain09} in the CS framework, allows us to provide sufficient conditions for exact support recovery. They are formally stated in the following lemma.

\newtheorem{Lem6}[Lem1]{Lemma}
\begin{Lem6} \label{MainLemma}
Let $\Pb_{\Abta^{\perp}} = \Ib - \Abta (\Abta^T \Abta)^{-1} \Abta^T$, 
and assume that $\zbh$ is the optimal solution to \eqref{fminlasso}. If
\begin{equation} \label{IrCoNoisy}
\Big \| \Abtac^T \big [ \sigma \Abta (\Abta^T \Abta)^{-1} \ubta +  (\lambda n)^{-1} \Pb_{\Abta^{\perp}} \wbt \big ] \Big \|_{\infty} \leq 1, 
\end{equation}
where $\ub$ is the associated subgradient of $\sum_i \fsig(|z_i|)$ at $\zbh$, then $\supp(\zbh) \subseteq \supp(\zbs)$. Moreover, if in addition to \eqref{IrCoNoisy},
\begin{equation} \label{noisySgnCond}
\Big | (\Abta^T \Abta)^{-1} \Big [ \Abta^T \wbt -\lambda n \sign(\zbsta) \odot f'(\frac{|\zbhta|}{\sigma}) \Big ] \Big | < |\zbsta|
\end{equation}
holds, 
then $\sign(\zbh) = \sign(\zbs)$.

\end{Lem6}

\smallskip

The first condition in the above lemma ensures that there is no false change point recognized by our algorithm, and the second one together with the first one guarantees $\sign(\zbh) = \sign(\zbs)$ which is stronger than what we are interested in; i.e., $\supp(\zbh) = \supp(\zbs)$. The results of this lemma have some connections to those obtained in \cite{CandP09,Wain09} in the framework of compressive sensing. More specifically, Lemma \ref{MainLemma} extends similar sufficient conditions for the $\ell_1$ penalty to the class of nonconvex penalties defined in Property \ref{Prop1}. However, this extension involves following a different approach to prove Lemma \ref{MainLemma}.

Assume that $\wbt \to \zerb$, which can be fulfilled by having $n \to \infty$ and $\lambda$ chosen carefully. Further, let $f(x) = 1 - e^x$. Then the sufficient conditions in Lemma \ref{MainLemma} simplify to
\begin{IEEEeqnarray*}{rCl}
& & \Big \| \Abtac^T \Abta (\Abta^T \Abta)^{-1} \big ( \sign(\zbsta) \odot e^{-|\zbhta|/\sigma} \big ) \Big \|_{\infty} < 1 \\
& & \Big | \lambda n (\Abta^T \Abta)^{-1} \big ( \sign(\zbsta) \odot e^{-|\zbhta|/\sigma} \big ) \Big | < |\zbsta|.
\end{IEEEeqnarray*}
In comparison to the associated conditions for \eqref{l1min}, where $e^{-|\zbhta|/\sigma}$ is replaced with the vector of ones, the above conditions are much easier to be satisfied. In fact, they 
show that when the magnitudes of the components of $\zbhta$ increase, the gradient vector (i.e., $\sign(\zbsta) \odot e^{-|\zbhta|/\sigma}$) will decrease exponentially in $\zbhta$, and we expect that the proposed approach detects the correct support easier than \eqref{l1min}. However, since the gradient vector depends on the solution of \eqref{fminlasso}, it is not possible to predict the performance improvement explicitly. Mathematically speaking, the above statement can be put in a probabilistic approach leading to the theorem below.

\newtheorem{Thm2}[Thm1]{Theorem}
\begin{Thm2} \label{MainThm}
Assume that
\begin{equation} \label{irrepCond}
\Big \| \Abtac^T \Abta (\Abta^T \Abta)^{-1} \big(\sign(\zbsta) \odot f'(|\zbhta|/\sigma) \big) \Big \|_{\infty} \leq 1 - \gamma
\end{equation}
for some $\gamma \in (0,1)$. Let $\alpha = \| f'(|\zbhta| / \sigma) \|_{\infty}$ and $\widetilde{s}_{\min}$ denote the smallest eigenvalue of $\Abta^T \Abta$. If
\begin{equation*}
\lambda > \frac{1}{\gamma \sqrt{2}} \sqrt{\frac{\ln n}{n} \sigma_w^2} = \lambda_0
\end{equation*}
and
\begin{equation} \label{zminInEq}
z^*_{\min} = \min_{i \in \tau} |z_i^*| > \lambda \Big (2 \sigma_w \sqrt{\frac{n}{\widetilde{s}_{\min}}} + n \| (\Abta^T \Abta)^{-1} \|_{\infty} \alpha \Big),
\end{equation}
then $\sign(\zbh) = \sign(\zbs)$ with a probability exceeding $P_1 \cdot P_2$, where
\begin{equation*}
P_1 = 1 - 2 \exp\big(-2 \frac{\gamma^2}{\sigma_w^2} (\lambda^2 - \lambda_0^2) n\big)
\end{equation*}
and $P_2 = 1 - 2 \exp\big(\ln(| \tau |) - 2 \lambda^2 n\big)$.

\end{Thm2}

\smallskip

Condition \eqref{irrepCond} in Theorem \ref{MainThm} is analogous to the well-known irrepresentable condition in \cite{ZhaoY06,CandP09,Wain09} which ensures correct support recovery for the lasso estimator \cite{BibkRA09}. Using the lasso equivalent form of $\ell_1$ mean filtering program introduced in \cite[Lem.~2.4]{RojaW14}, the irrepresentable condition for this program is
\begin{equation} \label{irrepCondl1}
\Big \| \Abtac^T \Abta (\Abta^T \Abta)^{-1} \sign(\zbsta) \Big \|_{\infty} \leq 1 - \gamma.
\end{equation}
To clarify how the result of Theorem \ref{MainThm} is compared to that of $\ell_1$ mean filtering, we should state the following proposition.

\smallskip

\newtheorem{Pro4}[Pro1]{Proposition}
\begin{Pro4} \label{l1Inf}
Assume that $\Bb = \Abtac^T \Abta (\Abta^T \Abta)^{-1}$, $\sbb$ denotes a sign vector consisting of components taking values of $\pm 1$, and $\tb$ denotes a weight vector in which $\forall i,\,0 < t_i \leq 1$. If for some $\sbb$, $\| \Bb \sbb \|_{\infty} = 1$, then
\begin{itemize}
  \item for every $\tb$ such that $\forall i,\,0 < t_i \leq 1$, one has\\*$\| \Bb ( \sbb \odot \tb) \|_{\infty} \leq 1$, and
  \item for every $\tb$ such that $\forall i,\,0 < t_i < 1$, one has\\*$\| \Bb ( \sbb \odot \tb) \|_{\infty} < 1$.
\end{itemize}
 Moreover, if for some $\sbb$, $\| \Bb \sbb \|_{\infty} < 1$, then for every $\tb$, we have $\| \Bb ( \sbb \odot \tb) \|_{\infty} < 1$.
\end{Pro4}

\smallskip

As shown above in \eqref{irrepCondl1}, $\| \Bb \sign(\zbsta) \|_{\infty} < 1$ is a sufficient condition for \eqref{l1min} to have a solution with the support containing in $\tau$ (Note that $\gamma$ cannot be equal to 0.). The above proposition shows that our approach will find a subset of $\tau$ as the set of CPs under a weaker condition. More precisely, it is shown in \cite{RojaW14} that when $\| \Bb \sign(\zbsta) \|_{\infty} = 1$ which can occur when the sign of two consecutive components of $\zbsta$ are the same, \eqref{l1min} will find false CPs with a probability that does not vanish as $n \to \infty$. The above proposition shows that even in the aforementioned case, one can still hope to recover $\tau$ using our approach especially when $z^*_{\min}$ is relatively large. This is because if $z^*_{\min} > 0$, then $\| \Bb ( \sign(\zbsta) \odot f'(|\zbhta|/\sigma)  \|_{\infty} < 1$ showing that \eqref{irrepCond} holds for some $\gamma > 0$.

Apart from the improvement shown in Proposition \ref{l1Inf}, the smallest nonzero elements of $\zbs$ needs to be much smaller in comparison to $\ell_1$ mean filtering to guarantee exact CP recovery. 
The explanation of this improvement is the following. 
As discussed above, to obtain results similar to those of Theorem \ref{MainThm} for $\ell_1$ mean filtering algorithm, one just needs to replace $f'(|\zbhta|/\sigma)$ with a vector of ones and $\alpha$ with 1 in the statement of this theorem. Now, let $\widehat{z}_{\min} = \min_{i \in \tau} |\widehat{z}_i|$. In the right hand side of inequality \eqref{zminInEq}, the first term corresponds to the noise power and the second one is due to the bias of the estimator. While this term equals $\lambda n \| (\Abta^T \Abta)^{-1} \|_{\infty}$ for \eqref{l1min}, for our approach, it has also the coefficient $\alpha$ which is equal to $e^{-\widehat{z}_{\min}/\sigma}$ for $f(x) = 1 - e^{-x}$. This shows that when $\widehat{z}_{\min}$ is large, the bias term and the smallest jump amplitude sufficient for CP recovery can be significantly smaller for our method.


\section{Numerical Simulations} \label{sec:NS}
In this section, the performance of the proposed algorithm is empirically assessed and compared with that of $\ell_1$ mean filtering and the algorithm of \cite{SelePB15}.

As discussed earlier when two consecutive jumps in the PWC signal are in same direction, $\ell_1$ mean filtering may detect false CPs known as the `stair-casing' problem. In contrast, if the jumps are in opposite directions, $\ell_1$ mean filtering performs well in general. To save space and show the effectiveness of our algorithm in the stair-casing problem, all simulations are done with a PWC signal generated according to the rule
\begin{equation} \label{truesigdef}
x^*_i = \left\{
\begin{IEEEeqnarraybox}[][c]{l?s}
\IEEEstrut
a & $1 \leq i \leq 50$ \\
2a & $51 \leq i \leq 100$ \\
3a & $101 \leq i \leq 200$
\IEEEstrut
\end{IEEEeqnarraybox}
\right.,
\end{equation}
where $a$ denotes the amplitude of the jumps. To generate the noise, 
the $w_i$'s are drawn independently from a zero-mean, unit-variance Gaussian distribution ($\sigma_w^2 = 1$). 
Moreover, the regularization parameter for $\ell_1$ mean filtering and our algorithm is set to $\lambda = 4 \sqrt{\sigma_w^2/n}$. However, the regularization parameter in the method of \cite{SelePB15} equals $\lambda' = \lambda n$ since the data fidelity term $\| \yb - \xb\|^2$ has a coefficient $\frac{1}{2}$ instead of $\frac{1}{2n}$ in the optimization problem. Consequently, for this method, the regularization parameter is set to $4 \sqrt{n \sigma_w^2n}$. For our algorithm, $f(x) = 1 - \exp(x)$ is used, the scaling parameter is always fixed to $\sigma = 4 \lambda n$, and the stopping criterion is that the relative distance between two consecutive solutions is less than $10^{-4}$.

The algorithm of \cite{SelePB15} is run with the MATLAB code provided as a supplement to \cite{SelePB15} using the default settings. In addition, the arctangent and log functions which are introduced in \cite{SelePB15} as instances of the nonconvex penalties, are both used in our comparisons. 
Finally, it is worth mentioning that all simulations are performed in MATLAB 8.3 environment using an Intel Core i7, 2.1 GHz processor with 8 GB of RAM under Microsoft Windows 7 operating system.

\subsection{Experiments}

\begin{figure}[tb]
\centering
\includegraphics[width=0.49\textwidth]{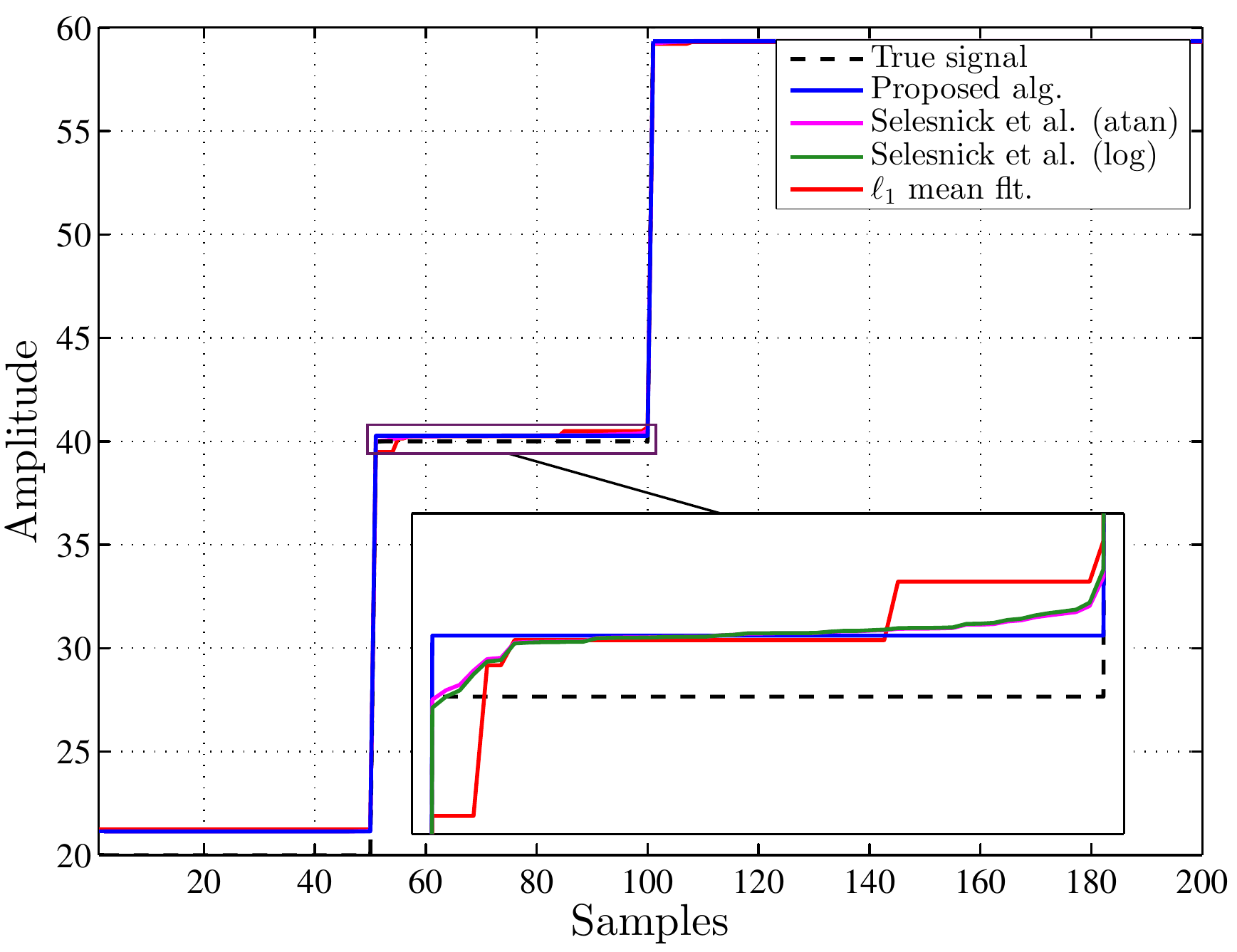}
\vspace{-0.5cm}
\caption{Estimations from application of $\ell_1$ mean filtering, the two instances of the algorithm of \cite{SelePB15}, and our algorithm are plotted for a single realization to show the stair-casing problem. The green and magenta curves show the results for the algorithm \cite{SelePB15} with the log and arctangent penalties, respectively. The true signal is generated according to \eqref{truesigdef}.} \label{fig:SimPlot}
\end{figure}

\begin{figure}[tb]
\centering
\includegraphics[width=0.49\textwidth]{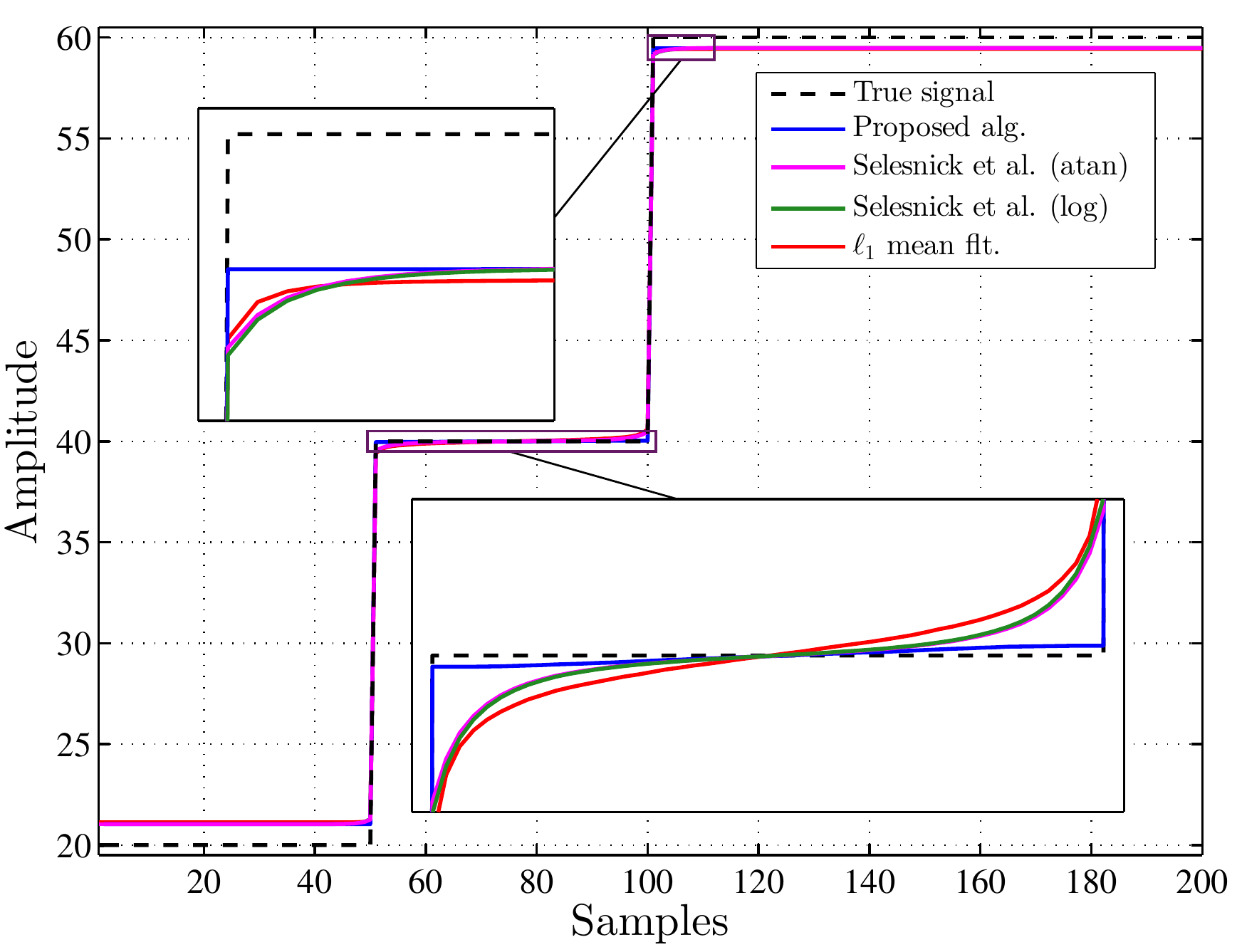}
\vspace{-0.5cm}
\caption{Averaged estimations from application of $\ell_1$ mean filtering, the algorithm of \cite{SelePB15}, and our algorithm with 10,000 MC realizations. Two instances of the algorithm \cite{SelePB15} have a very similar performance, and their curves are almost coincident.
} \label{fig:AvgPlot}
\end{figure}

\emph{Experiment 1.} 
To illustrate the stair-casing problem and effectiveness of our algorithm in resolving this issue, the true signal is generated with a jump amplitude $a = 20$. The three algorithms are applied, and the results are shown in Fig.~\ref{fig:SimPlot}. As can be seen, while $\ell_1$ mean filtering and the two instances of the algorithms of \cite{SelePB15} finds false jumps in the interval $50 < i < 100$, our proposed algorithm correctly identifies the CPs. Fig.~\ref{fig:AvgPlot} also illustrates the estimated solutions averaged over 10,000 Monte-Carlo (MC) realizations. It is remarkable that while our approach denoises the constant pieces better than others, it also follows the true signal with sharper edges. These are exactly the desired goals in TV denoising, and there is usually a trade off between them. So far as the complexity of the algorithms is concerned, the average computation times of $\ell_1$ mean filtering, the algorithms of \cite{SelePB15} with the arctangent and log penalties, and our approach are 0.7 ms, 2.8 ms, 2.8 ms, and 1.8 ms, respectively.

\begin{figure}[tb]
\centering
\includegraphics[width=0.49\textwidth]{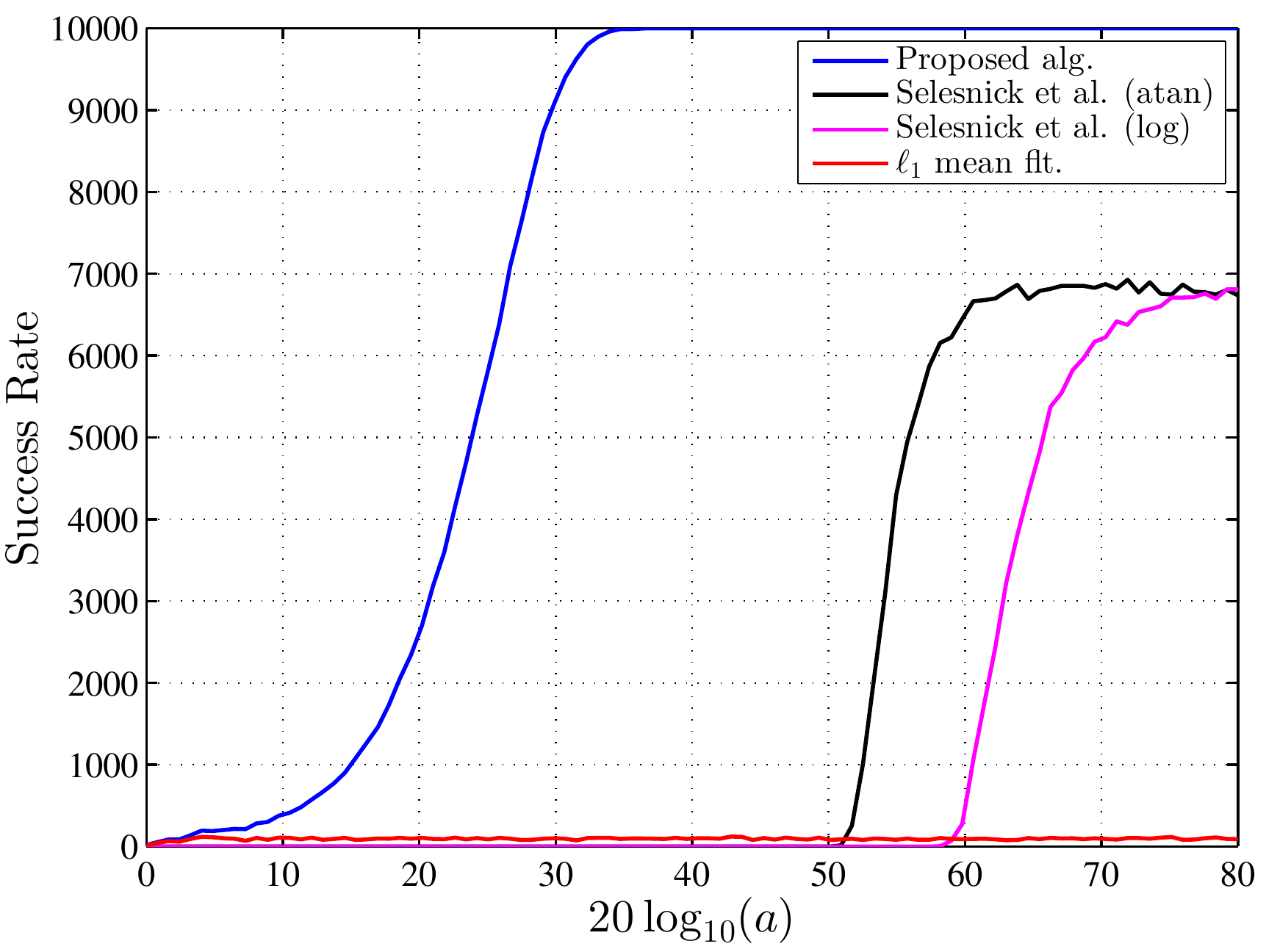}
\vspace{-0.5cm}
\caption{Success rate in recovering the CPs as a function of jump amplitude $a$ is plotted for $\ell_1$ mean filtering, the algorithm of \cite{SelePB15}, and our algorithm. The success rate is calculated using 10,000 MC realizations for each value of $a$.} \label{fig:SuppDetRate}
\end{figure}

\emph{Experiment 2.} To better understand the behaviour of the algorithms in detecting the CPs, an empirical probability of CP recovery is calculated in this experiment. The empirical probability is found as a function of jump amplitude while the noise variance is kept fixed. To this end, 
$a$ is swept from $1$ to $10^{4}$ in a logarithmic scale with a total number of 100 points. We declare that all CPs are identified, if an algorithm can detect the positions of them exactly without introducing any false CP. The success rate is then calculated as the number of successful identifications normalized by 
the number of 10,000 MC realizations. The success rate curve for all algorithms is depicted in Fig.~\ref{fig:SuppDetRate}. As can be seen clearly, $\ell_1$ mean filtering is unable to recover the true support even when the jump amplitude reaches $10^4$. Moreover, the method of \cite{SelePB15} can only detect the CPs at a rate of 0.7 when $a$ exceeds $10^3$. Our algorithm, however, starts to recover the change points with a rate of 1 when $a$ passes 50. This suggests that while our algorithm is consistent in recovering the change points in this experiment, the two other competitors are not. It also confirms that when $\| \Bb \sign(\zbsta) \|_{\infty} = 1$, even when $z^*_{\min}$ goes to $\infty$, it is not possible to avoid false CP recovery when \eqref{l1min} is used.

\section{Conclusion} \label{sec:C}
The idea of using a certain class of nonconvex penalties to regularize sparsity more tightly than the $\ell_1$ norm, appeared previously in \cite{MaleBS14,MaleKBJR15}, was extended in this paper to the mean filtering problem. Particularly, we replaced the $\ell_1$ penalty in $\ell_1$ mean filtering algorithm with one of these nonconvex penalties and arrived at a new optimization program. As the mean filtering problem is determined, contrary to \cite{MaleBS14,MaleKBJR15}, we were able to preserve the convexity of the optimization program under some conditions and proposed an efficient method with a convergence guarantee to solve it. 
To evaluate our algorithm, we established performance guarantees for exact change point recovery. We also assessed our method numerically which showed considerable superiority over $\ell_1$ mean filtering and the method of \cite{SelePB15} in terms of CP recovery.

\appendix
\section{Proofs} \label{app}


First, a few notations which will be used in the proofs are introduced.

\emph{Further Notations}: For symmetric matrices $\Yb,\Zb$, $\Yb \succeq \Zb$ means $\Yb - \Zb$ is positive semidefinite. $p\{\cdot\}$ denotes the probability of the event described in the braces, and $E\{\cdot\}$ represents the expected value.

\subsection{Proof of Theorem \ref{ConvexityThm}}
Using the variable change $\xb = \Mb_{n} \zb$, where
\begin{equation} \label{MDef}
\Mb_{n} =
\begin{bmatrix}
1 & 0 & 0 & \cdots & 0 \\
1 & 1 & 0 & \cdots & 0 \\
\vdots & \vdots & \ddots  & \ddots & \vdots \\
1 & 1 & \cdots & 1 & 0 \\
1 & 1 & \cdots & 1 & 1
\end{bmatrix}
\end{equation}
is $n \times n$ and full rank, the cost function $\frac{1}{2n} \| \yb - \xb \|^2 + \lamsig \sum_{i = 1}^{n-1} \fsig( |x_{i + 1} - x_{i}|)$ will be equal to
\begin{equation*}
g(\zb) = \frac{1}{2n} \Big \| \yb - z_1 \oneb -
\begin{bmatrix}
\zerb^T \\
\Mb_{n-1} \\
\end{bmatrix}
\zbt \Big \|^2 + \lamsig \sum_{i = 1}^{n-1} \fsig( |\widetilde{z}_{i}|),
\end{equation*}
where $\zbt = (z_2, \cdots, z_n)^T$. 
Following the same line of argument as in \cite[Lem.~2.4]{RojaW14}, it can be shown that
\begin{IEEEeqnarray}{rCl} \label{Fz1}
g(\zb) & = & 0.5 \Big(z_1 - \frac{1}{n} \oneb^T \big ( \yb -
\begin{bmatrix}
\zerb^T \\
\Mb_{n-1} \\
\end{bmatrix} \zbt \big)  \Big)^2 + \frac{1}{2n} \| \ybt - \Ab \zbt \|^2 \nonumber \\
& &  + \lamsig \sum_{i = 1}^{n-1} \fsig( |\widetilde{z}_{i}|),
\end{IEEEeqnarray}
where $\ybt$ and $\Ab$ are defined in Proposition \ref{LassoEqv}. Since the first term in \eqref{Fz1} is strictly convex in $\zb$, to prove strict convexity of $g(\zb)$, it suffices to show that the remaining terms which are denoted as $F(\zbt)$ are convex in $\zbt$. Let us define
\begin{equation*}
\phi(\zbt) = \lamsig \sum_{i = 1}^{n-1} \fsig( \widetilde{z}_{i}) \quad \text{and} \quad h(\zbt)=\frac{1}{2n} \| \ybt - \Ab \zbt \|^2,
\end{equation*}
then $F(\zbt) = h(\zbt) + \phi(|\zbt|)$. Since $\nabla^2 h(\zbt) \succeq \frac{1}{n} s_{\min} \Ib$, 
we can write that, for any $\rb$ and $\sbb$,
\begin{equation} \label{InEq1}
h(\rb) \geq h(\sbb) + \langle \rb - \sbb, \nabla h(\sbb) \rangle + \frac{s_{\min}}{2n} \| \rb - \sbb \|^2.
\end{equation}
It is also known that
\begin{equation*}
\nabla^2 \phi(\zbt) = \frac{\lamsig}{\sigma^2}  \diag\big(f''(\frac{\widetilde{z}_1}{\sigma}),\cdots,f''(\frac{\widetilde{z}_{n-1}}{\sigma}) \big) \succeq -\frac{\lamsig}{\sigma^2} \mu \Ib
\end{equation*}
for any $\zbt \geq \zerb$; 
thus, for any $\rb,\sbb \geq \zerb$, it can be written that
\begin{equation} \label{InEq2}
\phi(\rb) \geq \phi(\sbb) + \langle \rb - \sbb, \nabla \phi(\sbb) \rangle - \frac{\lamsig}{2\sigma^2} \mu \| \rb - \sbb \|^2.
\end{equation}
Adding $|\cdot|$ to the argument of the function $\phi$ in \eqref{InEq2}, we can write that, for any $\rb,\sbb$,
\begin{equation} \label{InEq3}
\phi(|\rb|) \geq \phi(|\sbb|) + \langle |\rb| - |\sbb|, \nabla \phi(|\sbb|) \rangle - \frac{\lamsig}{2\sigma^2} \mu \big\| |\rb| - |\sbb| \big\|^2,
\end{equation}
where $\nabla \phi(|\sbb|)$ denotes the gradient of $\phi$ at the point $|\sbb|$. Applying $\big \| |\rb| - |\sbb| \big \| \leq \| \rb - \sbb \|$, \eqref{InEq3} resorts to
\begin{equation} \label{InEq4}
\phi(|\rb|) \geq \phi(|\sbb|) + \langle |\rb| - |\sbb|, \nabla \phi(|\sbb|) \rangle - \frac{\lamsig}{2\sigma^2} \mu \| \rb - \sbb \|^2.
\end{equation}
Putting \eqref{InEq1} and \eqref{InEq4} together, we arrive at
\ifCLASSOPTIONjournal
\begin{IEEEeqnarray}{rCl} \label{InEq5}
F(\rb) & \geq & F(\sbb) + \langle \rb - \sbb, \nabla h(\sbb) \rangle + \langle |\rb| - |\sbb|, \nabla \phi(|\sbb|) \rangle \nonumber \\
 & &  + \frac{1}{2} (\frac{s_{\min}}{n} - \frac{\lamsig}{\sigma^2} \mu) \| \rb - \sbb \|^2.
\end{IEEEeqnarray}
\else
\begin{equation} \label{InEq5}
F(\rb) \geq F(\sbb) + \langle \rb - \sbb, \nabla h(\sbb) \rangle + \langle |\rb| - |\sbb|, \nabla \phi(|\sbb|) \rangle \nonumber + \frac{1}{2} (\frac{s_{\min}}{n} - \frac{\lamsig}{\sigma^2} \mu) \| \rb - \sbb \|^2.
\end{equation}
\fi
Let us define $\ub = \theta \sbb + (1 - \theta) \rb$ for $0 < \theta <1$. Applying \eqref{InEq5} twice on $(\rb,\ub)$ and $(\sbb,\ub)$ yields
\begin{IEEEeqnarray*}{rCl}
F(\rb) & \geq & F(\ub) + \langle \rb - \ub, \nabla h(\ub) \rangle + \langle |\rb| - |\ub|, \nabla \phi(|\ub|) \rangle \\
 & & + \frac{1}{2} (\frac{s_{\min}}{n} - \frac{\lamsig}{\sigma^2} \mu) \| \rb - \ub \|^2,\\
F(\sbb) & \geq & F(\ub) + \langle \sbb - \ub, \nabla h(\ub) \rangle + \langle |\sbb| - |\ub|, \nabla \phi(|\ub|) \rangle \\
& & + \frac{1}{2} (\frac{s_{\min}}{n} - \frac{\lamsig}{\sigma^2} \mu) \| \sbb - \ub \|^2.\\
\end{IEEEeqnarray*}
Multiplying both sides of the above inequalities by $1-\theta$ and $\theta$, respectively, and adding them together leads to
\ifCLASSOPTIONjournal
\begin{IEEEeqnarray}{rCl}
\IEEEeqnarraymulticol{3}{l}{\hspace{-1.7em} (1-\theta) F(\rb) + \theta F(\sbb)}\nonumber \\
\hspace{-1em} & \geq & F(\ub) + \langle (1 - \theta) |\rb| + \theta |\sbb| - |\ub|, \nabla \phi (|\ub|) \rangle \nonumber\\
\hspace{-1em} & & \! \! + \frac{1}{2} (\frac{s_{\min}}{n} - \frac{\lamsig}{\sigma^2} \mu) \Big[ (1 - \theta) \| \rb - \ub \|^2 + \theta \| \sbb - \ub \|^2 \Big]. \label{InEq6}
\end{IEEEeqnarray}
\else
\begin{IEEEeqnarray}{rCl}
(1-\theta) F(\rb) + \theta F(\sbb) & \geq & F(\ub) + \langle (1 - \theta) |\rb| + \theta |\sbb| - |\ub|, \nabla \phi (|\ub|) \rangle \nonumber\\
 & & + \frac{1}{2} (\frac{s_{\min}}{n} - \frac{\lamsig}{\sigma^2} \mu) \Big[ (1 - \theta) \| \rb - \ub \|^2 + \theta \| \sbb - \ub \|^2 \Big]. \label{InEq6}
\end{IEEEeqnarray}
\fi
To complete the proof, we need to show that $\langle (1 - \theta) |\rb| + \theta |\sbb| - |\ub|, \nabla \phi (|\ub|) \rangle \geq 0$. Since $\nabla \phi (|\ub|) \geq \zerb$,\footnote{From Property 1-(d) and 1-(e), it can be verified that $f'(x) \geq 0$ for $x \geq 0$.} it is sufficient to show that $(1 - \theta) |\rb| + \theta |\sbb| - |\ub| \geq \zerb$, which is simply verified by
\begin{equation*}
(1 - \theta) |\rb| + \theta |\sbb| = |(1 - \theta) \rb| + | \theta \sbb| \geq | (1 - \theta) \rb + \theta \sbb | = | \ub |.
\end{equation*}
Consequently, it can be concluded that, for any $\rb , \sbb,~0 < \theta < 1$, we have
\begin{equation*}
F(\theta \sbb + (1 - \theta) \rb) < \theta F(\sbb) + (1 - \theta) F(\rb)
\end{equation*}
provided that $\sigma^2 \geq \frac{\lamsig n}{s_{\min}} \mu$. As the objective function in \eqref{fminlasso} equals $F(\zb)$, $\sigma^2 > \frac{\lamsig n}{s_{\min}} \mu$ implies that \eqref{fminlasso} is strictly convex. This completes the proof.\hfill\QED




\subsection{Proof of Proposition \ref{InftyPro}}
First, it is shown that
\begin{equation*}
\lim_{\sigma \to \infty} \| \zbhs \| / \sigma  = 0.
\end{equation*}
Let us denote the objective function in \eqref{fminlasso} as $h(\zb)$. Optimality of $\zbhs$ implies that $h(\zbhs) \leq h(\zbt)$. Equivalently, we have
\begin{IEEEeqnarray}{rCl}
\frac{1}{2n \lambda \sigma} \| \ybt - \Ab \zbhs \|^2 & \leq & \sum_{i} \fsig(|\zt_i|) - \sum_{i} \fsig(|[\zbhs]_i|) \nonumber \\
&& + \frac{1}{2n \lambda \sigma} \| \ybt - \Ab \zbt \|^2 \nonumber\\
& \leq & \sum_{i} \fsig(|\zt_i|) + \frac{1}{2n \lambda \sigma} C, \label{InEq7}
\end{IEEEeqnarray}
where $C$ is an upper bound for $\| \ybt - \Ab \zbt \|^2$. Since $\zbt$ is bounded, Property \ref{Prop1}-(d) implies that
\begin{equation*}
\Big (\sum_{i} \fsig(|\zt_i|) = \sum_{i} f(|\zt_i|/\sigma) \Big) \to 0
\end{equation*}
when $\sigma \to \infty$. Hence, from inequality \eqref{InEq7}, we conclude that
\begin{equation} \label{LimLem10}
\lim_{\sigma \to \infty} \frac{\|\ybt - \Ab \zbhs\|}{\sqrt{\sigma}} = 0.
\end{equation}
As $\Ab$ is full column rank, \eqref{LimLem10} implies
\begin{equation}\label{BndNss}
\lim_{\sigma \to \infty} \frac{\|\zbhs\|}{\sqrt{\sigma}} = 0,
\end{equation}
which is stronger than what we need.

The Taylor expansion of $f(x)$ about $0$, for any $x \geq 0$, is equal to $f(x) = x + g(x)$, where
\begin{equation} \label{glim}
\lim_{x \to 0^{+}} \frac{g(x)}{x} = 0.
\end{equation}
Applying this expansion, we get
\begin{equation}\label{Texp}
\sum_i \fsig(|z_i|) = \sum_i f(|z_i|/\sigma) = \frac{1}{\sigma} \| \zb \|_1 + \sum_i g(|z_i|/\sigma).
\end{equation}
Substituting \eqref{Texp} in $h(\zbhs) \leq h(\zbt)$, we arrive at
\begin{IEEEeqnarray}{rCl}
\lambda \big (\| \zbhs \|_1 - \| \zbt \|_1 \big ) & \leq & (2n)^{-1} \big( \| \ybt - \Ab \zbt \|^2 - \| \ybt - \Ab \zbhs \|^2 \big) \nonumber \\
& & + \lambda \sigma \sum_{i \in \widetilde{\tau}} \frac{|\zt_i|}{\sigma} \frac{g(|\zt_i|/\sigma)}{|\zt_i|/\sigma} \nonumber \\
& & - \lambda \sigma \sum_{i \in \widehat{\tau}} \frac{\big |[\zbhs]_i \big |}{\sigma} \frac{g(\big |[\zbhs]_i \big |/\sigma)}{\big |[\zbhs]_i \big|/\sigma} \nonumber \\
& \overset{(a)}{\leq} & (2n)^{-1} \big( \| \ybt - \Ab \zbt \|^2 - \| \ybt - \Ab \zbhs \|^2 \big) \nonumber \\
& & + \lambda \| \zbt \|_1 \sum_{i \in \widetilde{\tau}} \frac{\Big |g(|\zt_i|/\sigma) \Big    |}{|\zt_i|/\sigma} \nonumber \\
& & + \lambda \| \zbhs \|_1 \sum_{i \in \widehat{\tau}}  \frac{\Big |g(\big |[\zbhs]_i \big |/\sigma)\Big |}{\big |[\zbhs]_i \big |/\sigma}, \label{l1Infty}
\end{IEEEeqnarray}
where $\widetilde{\tau}$ and $\widehat{\tau}$ designate the support sets of $\zbt$ and $\zbhs$, respectively. Moreover, for $(a)$, we used the inequality $\sum x_i y_i \leq \big(\sum |x_i|\big) \big(\sum |y_i|\big)$. \eqref{l1Infty} can be rearranged to
\ifCLASSOPTIONjournal
\begin{IEEEeqnarray}{rCl}
\IEEEeqnarraymulticol{3}{l}{\hspace{-1.7em} \lambda \| \zbhs \|_1 \Big [1 - \sum_{i \in \widehat{\tau}} \frac{\Big |g(\big|[\zbhs]_i \big |/\sigma)\Big|}{\big |[\zbhs]_i \big |/\sigma} \Big ]} \nonumber \\ \qquad \qquad
& \leq & (2n)^{-1} \big( \| \ybt - \Ab \zbt \|^2 - \| \ybt - \Ab \zbhs \|^2 \big) \nonumber \\
& & + \lambda \| \zbt \|_1 \Big [1 + \sum_{i \in \widetilde{\tau}} \frac{\big |g(|\zt_i|/\sigma) \big|}{|\zt_i|/\sigma} \Big ]. \label{finInEq}
\end{IEEEeqnarray}
\else
\begin{IEEEeqnarray}{rCl}
\lambda \| \zbhs \|_1 \Big [1 - \sum_{i \in \widehat{\tau}} \frac{\Big |g(\big|[\zbhs]_i \big |/\sigma)\Big|}{\big |[\zbhs]_i \big |/\sigma} \Big ] & \leq & (2n)^{-1} \big( \| \ybt - \Ab \zbt \|^2 - \| \ybt - \Ab \zbhs \|^2 \big) \nonumber \\
& & + \lambda \| \zbt \|_1 \Big [1 + \sum_{i \in \widetilde{\tau}} \frac{\big |g(|\zt_i|/\sigma) \big|}{|\zt_i|/\sigma} \Big ]. \label{finInEq}
\end{IEEEeqnarray}
\fi
Relations \eqref{BndNss} and \eqref{glim} show that
\begin{IEEEeqnarray}{rll}
\lim_{\sigma \to \infty} & \frac{\Big | g(\big|[\zbhs]_i \big |/\sigma) \Big |}{\big |[\zbhs]_i \big|/\sigma} & = 0 \label{InEqInfty1}\\
\lim_{\sigma \to \infty} & \frac{\big | g(|\zt_i|/\sigma) \big |}{|\zt_i   |/\sigma} & = 0 \label{InEqInfty2}.
\end{IEEEeqnarray}
Application of \eqref{InEqInfty1} and \eqref{InEqInfty2} on \eqref{finInEq} when $\sigma \to \infty$ leads to
\begin{equation*}
\lim_{\sigma \to \infty} \Big [ \lambda \| \zbhs \|_1 + \frac{1}{2n} \| \ybt - \Ab \zbhs \|^2 \Big ] \leq  \lambda \| \zbt \|_1 + \frac{1}{2n} \| \ybt - \Ab \zbt \|^2
\end{equation*}
which confirms that $\lim_{\sigma \to \infty} \zbhs = \zbt$ as $\zbt$ is unique.\hfill\QED

\subsection{Proof of Lemma \ref{MainLemma}}
Let $\zbtta$ denote the optimal solution to the restricted program \eqref{fminlasso_res}. We show that $\zbt$, generated from $\zbtta$ by setting the components in $\tau^{c}$ equal to 0, is the solution to the unrestricted program \eqref{fminlasso}, provided that \eqref{IrCoNoisy} holds. This confirms that the support set of the solution to \eqref{fminlasso} is a subset or equal to $\tau$.

As obtained in Lemma \ref{OCLemma}, the optimality condition for $\zbtta$ is
\begin{equation*} 
\frac{1}{n} \Abta^T ( \ybt - \Abta \zbtta) = \lamsig \ubtta,
\end{equation*}
where $\ubtta$ is a subgradient of $\sum \fsig(|z_i|)$ computed at $\zbtta$. Substituting $\ybt$ with $\Abta \zbsta + \wbt$, we obtain
\begin{equation} \label{OptCond_res}
\zbsta - \zbtta = (\Abta^T \Abta)^{-1} [\lamsig n \ubtta - \Abta^T \wbt ].
\end{equation}
To show that $\zbt$ is the solution to \eqref{fminlasso}, as Lemma \ref{OCLemma} suggests, it is sufficient to prove that
\begin{equation} \label{OptCond_unres}
\frac{1}{n} \Ab^T ( \ybt - \Ab \zbt) = \lamsig \ubt,
\end{equation}
where $\ubt = (\ubtta^T,\ubttac^T)^T$ and $\ubttac$ is the associated subgradient at $\zbttac$ satisfying $\| \ubttac \|_{\infty} \leq 1 / \sigma$. To do so, we can first write
\begin{IEEEeqnarray*}{rCl}
\ybt - \Ab \zbt & = & \Abta(\zbsta - \zbhta) + \wbt \\
                & = & \Abta (\Abta^T \Abta)^{-1} [\lamsig n \ubtta - \Abta^T \wbt ] + \wbt \\
                & = & \Pb_{\Abta^{\perp}} \wbt + \lamsig n \Abta (\Abta^T \Abta)^{-1} \ubtta.
\end{IEEEeqnarray*}
Consequently,
\begin{IEEEeqnarray}{lCl} \label{OC_Checked1}
\hspace{-2em}\Abta^T (\ybt - \Ab \zbt)  & = & \lamsig n \ubtta,\\
\hspace{-2em}\Abtac^T (\ybt - \Ab \zbt) & = & \Abtac^T \big [\Pb_{\Abta^{\perp}} \wbt + \lamsig n \Abta (\Abta^T \Abta)^{-1} \ubtta\big ].\label{OC_Checked2}
\end{IEEEeqnarray}
On the other hand, the optimality condition in \eqref{OptCond_unres} can be read as
\begin{equation*}
\frac{1}{n}
\begin{bmatrix}
\Abta^T \\
\Abtac^T
\end{bmatrix} ( \ybt - \Abta \zbtta) =
\frac{1}{n}
\begin{bmatrix}
\Abta^T ( \ybt - \Abta \zbtta)\\
\Abtac^T ( \ybt - \Abta \zbtta)\\
\end{bmatrix}
= \lamsig
\begin{bmatrix}
\ubtta \\
\ubttac \\
\end{bmatrix}.
\end{equation*}
Therefore, \eqref{OC_Checked1} shows that $\zbt$ satisfies the top block of the above optimality condition, and \eqref{OC_Checked2} together with \eqref{IrCoNoisy} justifies that $\ubttac = \Abtac^T \big [ (\lamsig n)^{-1} \Pb_{\Abta^{\perp}} \wbt + \Abta (\Abta^T \Abta)^{-1} \ubta \big ]$ is the valid subgradient vector since $\| \ubttac \|_{\infty} \leq 1 / \sigma$ and it is associated with the zero subvector $\zbttac$. This confirms that $\zbt = \zbh$ and $\supp(\zbh) \subseteq \supp(\zbs)$. To prove the second part, first notice that, from \eqref{OptCond_res}, we get
\begin{equation*}
\zbhta = \zbsta + (\Abta^T \Abta)^{-1} [\Abta^T \wbt - \lamsig n \ubta],
\end{equation*}
where $\zbt$ and $\ubtta$ are replaced by $\zbh$ and $\ubta$, respectively. Obviously, if
\begin{equation}\label{SigConsInter}
| \zbsta |  > \big | (\Abta^T \Abta)^{-1} [\Abta^T \wbt - \lamsig n \ubta] \big|,
\end{equation}
then $\sign(\widehat{z}_i) = \sign(z^*_i), \, \forall i \in \tau$ proving that $\sign(\zbh) = \sign(\zbs)$. On the hand, $\sign(\zbh) = \sign(\zbs)$ implies that $\ubta = \frac{1}{\sigma} \sign(\zbsta) \odot f'(\frac{|\zbhta|}{\sigma})$. Since given an optimal solution $\zbh$, the associated subgradient vector is unique (see the optimality condition \eqref{OC}), \eqref{noisySgnCond} is equivalent to \eqref{SigConsInter}. This completes the proof.\hfill\QED

\subsection{Two Auxiliary Lemmas}
To be able to prove Theorem \ref{MainThm}, we first need the following Lemmas.

\newtheorem{Lem4}[Lem1]{Lemma}
\begin{Lem4} \label{nrmBound}
Let $\ab_j$ denote the $j$th column of $\Ab$, then
\begin{equation*}
  \max_{1 \leq j \leq n - 1} \| \ab_j \|^2 \leq \frac{n}{4}.
\end{equation*}

\begin{IEEEproof}
It can be verified that $\| \ab_j \|^2 = j \frac{n - j}{n}$. If $n$ is even, $\| \ab_j \|^2$ is maximized with $j = \frac{n}{2}$, while it is maximized with $j = \frac{n}{2} - 0.5$ or $j = \frac{n}{2} + 0.5$, if $n$ is odd. Therefore, for even $n$, $\| \ab_j \|^2 \leq \frac{n}{4}$, and for odd $n$, $\| \ab_j \|^2 \leq \frac{1}{4}(n- 1) \frac{n + 1}{n} \leq \frac{n}{4}.$
\end{IEEEproof}
\end{Lem4}

\smallskip

\newtheorem{Lem5}[Lem1]{Lemma}
\begin{Lem5} \label{PrBounds}
For any $j \in \tau^c$,
\begin{equation*}
P\big \{ |\ab_j^T \Pb_{\Abta^{\perp}} \wbt | \geq t \big \} \leq 2 \exp(-\frac{2t^2}{n \sigma_w^2}),
\end{equation*}
where $t > 0$ is arbitrary. Moreover, for any $1 \leq j \leq | \tau |$,
\begin{equation*}
P\big \{ |\eb_j^T ( \Abta^T \Abta)^{-1} \Abta \wbt | \geq t \big \} \leq 2 \exp(-\frac{t^2 \widetilde{s}_{\min}}{2 \sigma_w^2}),
\end{equation*}
where $\widetilde{s}_{\min}$ and $\eb_j$ denote the smallest eigenvalue of $\Abta^T \Abta$ and the $j$th canonical basis vector of length $|\tau|$, respectively.

\begin{IEEEproof}
Let us define $U = \ab_j^T \Pb_{\Abta^{\perp}} \wbt$. $U$ is a zero-mean Gaussian random variable. To establish the claimed bound, we first need to find the variance of $U$ which depends on the variance of $\wbt$. Notice that unlike $\wb$, the components of $\wbt$ are not independent, and for $ i \neq k$, it is possible to write
\ifCLASSOPTIONjournal
\begin{IEEEeqnarray*}{rCl}
\IEEEeqnarraymulticol{3}{l}{ E \{\wt_i \wt_k \} }  \\ \qquad
& = & \frac{1}{n^2} E \Big \{ \Big [ (n - 1) w_i - \sum_{j \neq i} w_j \Big] \Big [(n - 1) w_k - \sum_{j \neq k} w_j \Big] \Big \},\\
& = & - \frac{1}{n ^ 2} \Big [ (n - 1) (E \{ w_i^2 \} + E \{ w_k^2 \}) - \sum_{\substack{j \neq k\\j \neq i}} E \{ w_j^2 \} \Big],\\
& = & - \frac{1}{n} \sigma_{w}^2.
\end{IEEEeqnarray*}
\else
\begin{IEEEeqnarray*}{rCl}
E \{\wt_i \wt_k \} & = & \frac{1}{n^2} E \Big \{ \Big [ (n - 1) w_i - \sum_{j \neq i} w_j \Big] \Big [(n - 1) w_k - \sum_{j \neq k} w_j \Big] \Big \},\\
& = & - \frac{1}{n ^ 2} \Big [ (n - 1) (E \{ w_i^2 \} + E \{ w_k^2 \}) - \sum_{\substack{j \neq k\\j \neq i}} E \{ w_j^2 \} \Big],\\
& = & - \frac{1}{n} \sigma_{w}^2.
\end{IEEEeqnarray*}
\fi
Also, it can be verified that
\begin{equation*}
E \{\wt_i^2 \} = (\frac{n - 1}{n})^2 \sigma_{w}^2 + \frac{n - 1}{n ^ 2} \sigma_{w}^2 = \frac{n - 1}{n} \sigma_{w}^2.
\end{equation*}
Now, let us define $\bb = \Pb_{\Abta^{\perp}} \ab_j$; it is possible to write
\begin{IEEEeqnarray*}{rCl}
E \{ U^2 \} & = & \sum_{i,k} b_i b_k E\{ \wt_i \wt_k \} \\
& = & \frac{n - 1}{n} \sigma_w^2 \sum_{i} b_i^2 - \frac{1}{n} \sigma_w^2 \sum_{\substack{i,k\\i \neq k}} b_i b_k, \\
& = & \| \bb \|^2 \sigma_w^2 - \frac{1}{n} \sigma_w^2 \Big (\sum_i b_i \Big)^2, \\
& \leq & \| \bb \|^2 \sigma_w^2 = \| \Pb_{\Abta^{\perp}} \ab_j \|^2 \sigma_w^2, \\
& \leq & \| \ab_j \|^2 \sigma_w^2, \\
& \overset{(a)}{\leq} &  \frac{n}{4} \sigma_w^2,
\end{IEEEeqnarray*}
where $(a)$ follows from Lemma \ref{nrmBound}. Using Chernoff's bound and optimizing it, we can get $P\{ |U| \geq t \} \leq 2 \exp(-\frac{2t^2}{n \sigma_w^2})$ completing the proof of the first part.

For the second part, let us define $\db = \Abta ( \Abta^T \Abta)^{-1} \eb_j$ and $V = \db^T \wbt$. Again, $V$ is a zero-mean Gaussian random variable, and to establish the claimed bound, we first need to find its variance. It can be started from
\begin{IEEEeqnarray*}{rCl}
 E \{ V^2 \} &  = & \sum_{i,k} d_i d_k E\{ \wt_i \wt_k \}, \\
 & = & \frac{\sigma_w^2}{n} \Big[ (n - 1) \sum_{i} d_i^2 - \sum_{\substack{i,k\\i \neq k}} d_i d_k \Big], \\
& = & \| \db \|^2 \sigma_w^2 - \frac{1}{n} \sigma_w^2 \Big (\sum_i d_i \Big)^2, \\
& \leq & \| \db \|^2 \sigma_w^2.
\end{IEEEeqnarray*}
On the other hand, we know that
\begin{equation*}
\eb_j^T ( \Abta^T \Abta)^{-1} \eb_j \leq \frac{1}{\widetilde{s}_{\min}};
\end{equation*}
thus, $E\{ V^2 \} \leq \sigma_w^2 / \widetilde{s}_{\min} $. Following the same line of argument as in the proof of the first part, we get $P\{ |\eb_j^T ( \Abta^T \Abta)^{-1} \Abta \wbt | \geq t \} \leq 2 \exp(-\frac{t^2 \widetilde{s}_{\min}}{2 \sigma_w^2})$.
\end{IEEEproof}

\end{Lem5}

\subsection{Proof of Theorem \ref{MainThm}}

The proof of Theorem \ref{MainThm} is inspired by the proof of \cite[Thm.~1]{Wain09}.
We start with checking the first condition in Lemma \ref{MainLemma}. As discussed in the proof of Lemma \ref{MainLemma}, under the assumptions made for the optimal solution $\zbh$ in this theorem or in Lemma \ref{MainLemma}, the subgradient vector $\ub$ is unique, and the only possible choice for $\ubta$ is $\ubta = \frac{1}{\sigma}\sign(\zbsta) \odot f'(|\zbhta| / \sigma)$.
Now, let us denote
\begin{equation*}
S = \Big \| \Abtac^T \big [ \sigma \Abta ( \Abta^T \Abta)^{-1} \ubta + \frac{1}{\lambda n} \Pb_{\Abta^{\perp}} \wbt \big ] \Big \|_{\infty};
\end{equation*}
we can write that
\begin{IEEEeqnarray}{rCl}
S & \leq & \Big \| \Abtac^T \Abta (\Abta^T \Abta)^{-1} \big ( \sign(\zbsta) \odot f'(|\zbhta|/\sigma) \big) \Big \|_{\infty} \nonumber \\
& & + \frac{1}{\lambda n} \| \Abtac^T \Pb_{\Abta^{\perp}} \wbt \|_{\infty} \nonumber \\
& \leq & (1 - \gamma)  + \frac{1}{\lambda n} \| \Abtac^T \Pb_{\Abta^{\perp}} \wbt \|_{\infty}. \label{InEqThm1}
\end{IEEEeqnarray}
Next, we try to find an upperbound for the probability that the second term in r.h.s. of \eqref{InEqThm1} is greater than or equal to $ \gamma$. Notice that
\begin{IEEEeqnarray*}{rCl}
P\Big\{ \frac{\| \Abtac^T \Pb_{\Abta^{\perp}} \wbt \|_{\infty}}{\lambda n} \geq \gamma  \Big\} & \overset{(a)}{\leq} & \sum_{j \in \tau^c} P\big\{|\ab_j^T \Pb_{\Abta^{\perp}} \wbt | \geq \lambda n \gamma \big\} \\
& \overset{(b)}{\leq} & 2 e^{\ln(n - 1 - |\tau|) -2 n\frac{\lambda^2  \gamma^2 }{\sigma_w^2}},
\end{IEEEeqnarray*}
where $(a)$ and $(b)$ follow from the union bound and Lemma \ref{PrBounds}, respectively. As a consequence of the above inequality, if one chooses $\lambda > \lambda_0$, then \eqref{IrCoNoisy} will hold with a probability larger than $P_1 = 1 - 2 \exp\big(-2 \frac{\gamma^2}{\sigma_w^2} (\lambda^2 - \lambda_0^2) n\big)$.

To fulfill \eqref{noisySgnCond} in Lemma \ref{MainLemma}, it is sufficient to have
\begin{equation*}
| z^*_{\min} |  \geq \Big \| (\Abta^T \Abta)^{-1} \big(\Abta^T \wbt - \lambda n \sign(\zbsta) \odot f'(\frac{|\zbh|}{\sigma}) \big) \Big \|_{\infty} = T.
\end{equation*}
For $T$, we have
\begin{equation*}
T \leq \| (\Abta^T \Abta)^{-1} \Abta^T \wbt \|_{\infty} + \lambda n  \| (\Abta^T \Abta)^{-1} \|_{\infty} \alpha.
\end{equation*}
Application of Lemma \ref{PrBounds} and the union bound lead to
\begin{equation*}
P\big\{ \| (\Abta^T \Abta)^{-1} \Abta^T \wbt \|_{\infty} \geq t \big\} \leq 2 \exp\big( \ln(|\tau|) - \frac{t^2 \widetilde{s}_{\min}}{2 \sigma_w^2} \big).
\end{equation*}
By choosing $t = 2 \sigma_w \sqrt{\frac{n}{\widetilde{s}_{\min}}} \lambda$,
\begin{equation*}
|z^*_{\min}| \geq \lambda (2 \sigma_w \sqrt{\frac{n}{\widetilde{s}_{\min}}} + n \| (\Abta^T \Abta)^{-1} \|_{\infty} \alpha)
\end{equation*}
is a sufficient condition to satisfy \eqref{noisySgnCond} with a probability larger than $1 - 2 \exp(\ln(| \tau |) - 2 \lambda^2 n)$.\hfill\QED

\subsection{Proof of Proposition \ref{l1Inf}}
A direct consequence of Lemma 2.6 in \cite{RojaW14} is as follows. In each row of $\Bb$, at most two components are nonzero and for every $i$ and $j$, $0 \leq B_{ij} < 1$. Moreover, when the number of nonzero components in a certain row is two, they are at two consecutive positions and sum to 1. This implies that for every sign vector $\sbb$, $\| \Bb \sbb \|_{\infty} \leq 1$. Furthermore, $\| \Bb \sbb \|_{\infty} = 1$ whenever there is a same sign pattern at two consecutive components of $\sbb$ corresponding to two nonzero components in some row of $\Bb$. In this case, it is clear that for any weight vector $\tb$, $\| \Bb (\sbb \odot \tb) \|_{\infty} \leq 1$; however, if $\forall i,~0 < t_i < 1$, we have $\| \Bb (\sbb \odot \tb) \|_{\infty} < 1$. The second part of the claim relates to the case that $\sbb$ is chosen, if it is possible, such that for every row of $\Bb$ with two nonzero components, the associated sign elements of $\sbb$ have opposite values. Obviously, in this case, $\| \Bb \sbb \|_{\infty} < 1$ and $\| \Bb (\sbb \odot \tb) \|_{\infty} < 1$.\hfill\QED

\bibliographystyle{IEEEbib}
\bibliography{IEEEabrv,MeanFiltering}

\end{document}